\documentclass[12pt]{article}
\usepackage{amssymb,amsmath,epsfig}
\allowdisplaybreaks

\begin{document}
\title{\bf Noether Symmetries and Anisotropic Universe in Energy-Momentum Squared Gravity}
\author{M. Sharif \thanks {msharif.math@pu.edu.pk} and M. Zeeshan Gul
\thanks{mzeeshangul.math@gmail.com}\\
Department of Mathematics, University of the Punjab,\\
Quaid-e-Azam Campus, Lahore-54590, Pakistan.}

\date{}
\maketitle

\begin{abstract}
This paper explores exact cosmological solutions of anisotropic
universe model through Noether symmetry technique in energy-momentum
squared gravity. This theory resolves the primordial singularity and
provides viable cosmological consequences in the early universe. We
consider specific models of this theory and evaluate Noether
equations, symmetry generators and corresponding conserved
parameters. We then find exact solutions through conserved
parameters and analyze their graphical behavior for different
cosmological parameters. It is found that the behavior of these
parameters is consistent with recent observations indicating that
this theory supports current cosmic accelerated expansion. We
conclude that conserved parameters are very helpful to obtain exact
cosmological solutions.
\end{abstract}

\section{Introduction}

The current cosmic acceleration has been the most dazzling and
interesting outcome for cosmologists over the last two decades.
Several cosmological observations such as supernovae type Ia, cosmic
microwave background (CMB), Planck data, etc., justify that the
universe is in the accelerated expansion phase \cite{1}. Scientists
believe that this expansion is the outcome of an enigmatic force
known as dark energy (DE) which has large negative pressure. This
ambiguous force has inspired many scientists to uncover its hidden
aspects. Einstein introduced cosmological constant $(\Lambda)$ in
his field equations to explain the mysterious aspects of DE which is
known as the cosmological constant cold dark matter ($\Lambda$CDM)
model. But, this approach has two main issues such as fine-tuning
and coincidence problems. The first issue arises because of the
large difference between observed and predicted values of energy
density whereas the second problem describes that if the energy
densities of DE and dark matter (DM) are identical then why current
cosmic acceleration is observed? \cite{2}.

Several modifications of general relativity (GR) have proposed
different approaches to solve these problems dubbed as modified
theories. These proposals are assumed as the most elegant and
significant approaches to reveal the dark universe. These theories
are developed by modifying the geometric and matter parts of the
Einstein-Hilbert (EH) action. The strong relation between gravity
and relativistic objects at cosmological scales indicates that
alternative gravitational theories should be established from the
curvature part of the EH action. The simplest modification of GR is
$f(\mathcal{R})$ gravity which is constructed by substituting the
generic function of Ricci scalar $(\mathcal{R})$ in the curvature
part of the EH action. To understand the viability of this theory,
detailed literature has been made available in \cite{3}.

The $f(\mathcal{R})$ theory has further been modified by including
some interactions between geometric and matter parts. These
interactions describe the rotation curves of galaxies and different
cosmic evolutionary eras. An additional force appears in this
framework due to the non-conserved stress-energy tensor and results
in the non-geodesic motion of the particles. These interaction
proposals are extremely helpful to comprehend mysterious aspects of
the cosmos. The coupling of matter-Lagrangian
$(\mathrm{L}_{\mathrm{M}})$ and curvature has been introduced in
\cite{4} named $f(\mathcal{R}, \mathrm{L}_{\mathrm{M}})$ theory. The
non-minimal interection between geometric and matter sectors was
formulated in \cite{5} named $f(\mathcal{R},
\mathrm{T},\mathcal{R}_{\mu\nu} \mathrm{T}^{\mu\nu})$ gravity, where
the stress-energy tensor is denoted by $\mathrm{T}$.

The presence of singularities is considered the most critical issue
in GR as these are predicted at large regimes, where GR is invalid
due to quantum impacts. In this perspective, a new generalization to
GR has been established by adding a non-linear term
$(\mathbf{T}^{2}=\mathrm{T}^{\mu\nu}\mathrm{T}_{\mu\nu})$ in the
generic action named $f(\mathcal{R}, \mathbf{T}^{2})$ theory which
is also known as \textit{energy-momentum squared gravity} (EMSG)
\cite{6}. This theory provides a specific relation between curvature
and matter parts and contains an extra force
$(\mathrm{T}^{\mu\nu}\mathrm{T}_{\mu\nu})$ together with
$f(\mathcal{R})$ which gives a better description to expose the
mysterious universe. The field equations include squared and product
entities of fluid parameters that are helpful to describe various
cosmological results. This approach has a small-scale parameter
corresponding to finite maximum energy density. As a result, it has
a bounce in the early times and resolves primordial singularity. It
is worth mentioning here that this theory describes the complete
cosmic history including the inflationary era as well as cosmic
evolution similar to that in the $\Lambda$CDM model.

The analytic solutions of isotropic spacetime with a specific EMSG
model have been discussed in \cite{7}. The physically viable compact
objects and hydrostatic equilibrium equations have been studied in
\cite{8}. Different coupling models of EMSG have been analyzed and
found that the proposed models can explain the current cosmic
acceleration \cite{9}. Barbar et al \cite{10} examined viability of
the bouncing universe in the same theory and explored that our
universe was non-singular in the early times. The geometry of
self-gravitating objects with quark matter and thermodynamic
characteristics of a black hole have been analyzed in \cite{11}.
Recently, we have examined the dynamics of spherical as well as
cylindrical collapse with various matter distributions in this
framework and found that modified terms of EMSG reduces the collapse
rate \cite{12}. The preceding literature demonstrates that this
theory requires more consideration in different contexts.

Many cosmic observations such as Wilkinson Microwave Anisotropy
Probe, Planck satellites and CMB indicate that the current cosmos is
isotropic and homogeneous at large scales. This cosmic stage is
determined by the Friedmann-Robertson-Walker (FRW) spacetime, which
ignores the anisotropy as well as all cosmic structures. However,
the universe was discovered to be spatially homogeneous and
anisotropic at the early times. The anisotropy is still analyzed in
the present cosmos as a CMB temperature. Bianchi type (BT) universe
models are considered the most significant and captivating models
which can explain the impact of anisotropy in the early times. These
anisotropic models demonstrate that the initial anisotropy affects
the fate of rapid expansion which will continue for large values of
anisotropy. If the initial anisotropy is minor then the rapid cosmic
expansion will stop leading to a highly isotropic cosmos \cite{13}.

Many researchers have studied these models from various
perspectives. Akarsu and Kilinc \cite{14} analyzed the BT-I model
with anisotropic fluid and found that the constant effective energy
density and equation of state (EoS) variable are responsible for
cosmic expansion. Yadav and Saha \cite{15} examined the BT-I
cosmological model with dominance of DE and found that DE leads to
the current cosmic accelerated expansion. Adhav \cite{16}
investigated exact cosmological solutions of anisotropic universe in
$f(\mathcal{R},\mathrm{T})$ gravity. Shamir \cite{18} discussed
exact solutions of the BT-I universe and analyzed their behavior
through different physical parameters in the same gravity. Sharif
and Jabbar \cite{19} studied the stability criteria of the BT-I
spacetime in $f(\mathcal{T})$ gravity, where $\mathcal{T}$ is the
torsion scalar.

Symmetry plays a crucial role in the study of cosmology and
gravitational physics. Accordingly, Noether symmetry technique is
considered the most efficient approach that describes a correlation
between symmetry generators and conserved parameters of a physical
system \cite{21}. This method is widely used to investigate exact
solutions in various aspects which are then discussed in terms of
cosmic features. Moreover, this technique gives a useful method to
establish conserved quantities. The conservation laws are most
important in analyzing various physical phenomena. These laws are
the specific cases of Noether theorem which states that every action
with a differentiable symmetry yields conservation law. This theorem
is significant as it gives a link among conserved parameters and
symmetries of a dynamical system \cite{22}. A lot of interesting
work has been examined in this framework \cite{23}.

Noether symmetries have several useful applications in alternative
gravitational theories. Capozziello et al \cite{24} found exact
cosmological solutions of spherical spacetimes through Noether
symmetry approach in $f(\mathcal{R})$ theory. Shamir et al \cite{25}
examined the stability of $f(\mathcal{R})$ models for static sphere
and FRW universe model through this approach. Kucukakca \cite{26}
formulated exact isotopic solutions through Noether symmetry in the
scalar-tensor theory. Sharif and Waheed \cite{27} investigated BT-I
and FRW spacetimes through Noether symmetries in the same theory.
Sharif and Shafique \cite{27a} investigated the BT-I spacetime
through Noether symmetries in scalar-field gravity. Momeni et al.
\cite{28} analyzed exact cosmological solutions through Noether
symmetries in $f(\mathcal{R},\mathrm{T})$ theory.

Sharif and Fatima \cite{29} derived exact cosmological solutions of
the FRW universe by Noether symmetries in $f(\mathrm{G})$ theory,
where $G$ is the Gauss-Bonnet term and examined the cosmic
acceleration in terms of the scale parameter. Shamir and Ahmad
\cite{30} examined various cosmological solutions with different
fluid configurations in $f(\mathrm{G}, \mathrm{T})$ gravity.
Bahamonde et al \cite{32} used this technique to examine new exact
spherical solutions in $f(\mathcal{R},\phi,\chi)$ gravity, where
$\phi$ is a scalar field and $\chi$ defines the kinetic term of
$\phi$. Bahamonde et al \cite{33} used Noether symmetry technique to
obtain wormhole solutions in $f(\mathcal{T})$ theory. Recently, we
have obtained exact cosmological solutions in
$f(\mathcal{R},\mathbf{T}^{2})$ gravity and analyzed their behavior
through various physical quantities \cite{34}. We have also
investigated wormhole solutions and geometry of compact stellar
objects in this background \cite{35}.

This paper examines Noether symmetries for BT-I universe in the
context of EMSG. We determine symmetry generators with corresponding
conserved parameters to find exact cosmological solutions for
different EMSG models and examine their graphical behavior through
different physical parameters. The plan of the article is as
follows. Section \textbf{2} investigates basic formalism of EMSG.
Section \textbf{3} provides a detailed analysis of Noether
symmetries. In section \textbf{4}, we derive exact BT-I solutions
using Noether symmetries and discuss them through graphs. We
summarize our results in the last section.

\section{Energy-Momentum Squared Gravity}

This section formulates the equations of motion with perfect matter
configuration in the framework of EMSG. The EH action of this
modified gravity is defined as \cite{6}
\begin{equation}\label{1}
S=\int \sqrt{-g}\left(\frac{f\left
(\mathcal{R},\mathbf{T}^{2}\right)}{2\kappa^2}+\mathrm{L}
_{\mathrm{M}}\right)d^4x,
\end{equation}
where  $\kappa^2$ is coupling constant. Here, the possibility of
exact solutions is increased than GR due to the presence of extra
degrees of freedom. It is assumed that some significant results will
be achieved to examine the current cosmic acceleration because of
the matter source. The following field equations are formulated by
varying the action corresponding to the metric tensor
\begin{equation}\label{2}
\mathcal{R}_{\mu\nu}f_{\mathcal{R}}+g_{\mu\nu}\Box
f_{\mathcal{R}}-\nabla_{\mu}
\nabla_{\nu}f_{\mathcal{R}}-\frac{1}{2}g_{\mu\nu}f=
\mathrm{T}_{\mu\nu}-\Theta_{\mu\nu}f_{\mathbf{T}^{2}},
\end{equation}
where $f\equiv f(\mathcal{R}, \mathbf{T}^{2})$,
$f_{\mathbf{T}^{2}}=\frac{\partial f}{\partial \mathbf{T}^{2}}$,
$f_{\mathcal{R}}= \frac{\partial f}{\partial \mathcal{R}}$,
$\Box=\nabla_{\mu}\nabla^{\mu}$ and
\begin{eqnarray}\label{3}
\Theta_{\mu\nu}
=-2\mathrm{L}_{\mathrm{M}}\left(\mathrm{T}_{\mu\nu}-\frac{1}{2}
g_{\mu\nu}
\mathrm{T}\right)-4\frac{\partial^{2}\mathrm{L}_{\mathrm{M}}}{\partial
g^{\mu\nu}\partial g^{\alpha\beta}}\mathrm{T}^{\alpha\beta}
-\mathrm{T}\mathrm{T}_{\mu\nu}+2\mathrm{T}_{\mu}^{\alpha}\mathrm{T}_{\nu\alpha}.
\end{eqnarray}
The energy-momentum tensor describes the matter configuration in
gravitational physics and gives dynamical variables with specific
physical characteristics. We consider isotropic fluid configuration
as
\begin{equation}\label{4}
\mathrm{T}_{\mu\nu}=
(\rho+\mathrm{p})\mathrm{U}_{\mu}\mathrm{U}_{\nu}+\mathrm{p}g_{\mu\nu},
\end{equation}
Rearranging Eq.(\ref{2}), we have
\begin{equation}\label{6}
\mathcal{G}_{\mu\nu}=
\frac{1}{f_{\mathcal{R}}}\left(\mathrm{T}_{\mu\nu}^{(\mathrm{D})}
+\mathrm{T}_{\mu\nu}\right)=\mathrm{T}_{\mu\nu}^{eff},
\end{equation}
where Einstein tensor is denoted by $\mathcal{G}_{\mu\nu}$ and
$\mathrm{T}_{\mu\nu}^{(D)}$ represents the additional effects of
EMSG that contain higher-order curvature terms due to the
modification in geometric part expressed as
\begin{equation}\label{7}
\mathrm{T}_{\mu\nu}^{(\mathrm{D})}= \frac{1}{2}g_{\mu\nu}
\left(f-\mathcal{R} f_{\mathcal{R}}\right)-g_{\mu\nu}\Box
f_{\mathcal{R}}+\nabla_{\mu}\nabla_{\nu}f_{\mathcal{R}}
-\Theta_{\mu\nu}f_{\mathbf{T}^{2}}.
\end{equation}
Equation (\ref{6}) determines that the dark source terms provide
matter contents of the universe. These additional terms contain all
the fluid elements that could be useful to reveal the dark
characteristics of the universe.

In order to analyze the anisotropic universe, we consider BT-I
universe model as
\begin{equation}\label{8}
ds^{2}= -d\mathrm{t}^{2}+\mathrm{a}^{2}(\mathrm{t})d\mathrm{x}^{2}
+\mathrm{b}^{2}(\mathrm{t})(d\mathrm{y}^{2}+d\mathrm{z}^{2}).
\end{equation}
The resulting equations of motion become
\begin{eqnarray}\nonumber
\rho^{eff}&=&\frac{1}{f_{\mathcal{R}}}\left[\rho-\frac{1}{2}f
+(3\mathrm{p}^{2}+\mathrm{\rho}^{2}+4\mathrm{p}\mathrm
{\rho})f_{\mathbf{T}^{2}}-(\dot{\mathrm{a}}\mathrm{a}^{-1}+2
\dot{\mathrm{b}}\mathrm{b}^{-1})\right.\\\label{9}&\times&\left.
(\dot{\mathcal{R}}f_{\mathcal{R}\mathcal{R}}+\dot{\mathbf{T}}^{2}
f_{\mathcal{R}\mathbf{T}^{2}})+\frac{1}{2}\mathcal{R}f_{\mathcal{R}}
\right],
\\\nonumber
\mathrm{p}^{eff}&=&\frac{1}{f_{\mathcal{R}}}\left[\mathrm{p}+\frac{1}
{2}(f-\mathcal{R}f_{\mathcal{R}})+2\dot{\mathrm{b}}\mathrm{b}^{-1}
(\dot{\mathcal{R}}f_{\mathcal{R}\mathcal{R}}+\dot{\mathbf{T}}^{2}
f_{\mathcal{R}\mathbf{T}^{2}})+\ddot{\mathcal{R}}f_{\mathcal{R}\mathcal{R}}
\right.\\\label{10}
&+&\left.\ddot{\mathbf{T}}^{2}f_{\mathcal{R}\mathbf{T}^{2}}
+\dot{\mathcal{R}}^{2}f_{\mathcal{R}\mathcal{R}\mathcal{R}}
+\dot{\mathbf{T}}^{2}f_{\mathcal{R}\mathbf{T}^{2}\mathbf{T}^{2}}
+2\dot{\mathcal{R}}\dot{\mathbf{T}}f_{\mathcal{R}\mathcal{R}
\mathbf{T}^{2}}\right],
\\\nonumber
\mathrm{p}^{eff}&=&\frac{1}{f_{\mathcal{R}}}\left[\mathrm{p}+\frac{1}
{2}(f-\mathcal{R}f_{\mathcal{R}})+(\dot{\mathrm{a}}\mathrm{a}^{-1}+\dot{\mathrm{b}}\mathrm{b}^{-1})
(\dot{\mathcal{R}}f_{\mathcal{R}\mathcal{R}}+\dot{\mathbf{T}}^{2}
f_{\mathcal{R}\mathbf{T}^{2}})+\ddot{\mathcal{R}}f_{\mathcal{R}\mathcal{R}}
\right.\\\label{10a}
&+&\left.\ddot{\mathbf{T}}^{2}f_{\mathcal{R}\mathbf{T}^{2}}
+\dot{\mathcal{R}}^{2}f_{\mathcal{R}\mathcal{R}\mathcal{R}}
+\dot{\mathbf{T}}^{2}f_{\mathcal{R}\mathbf{T}^{2}\mathbf{T}^{2}}
+2\dot{\mathcal{R}}\dot{\mathbf{T}}f_{\mathcal{R}\mathcal{R}
\mathbf{T}^{2}}\right],
\end{eqnarray}
where the dot is derivative corresponding to the temporal
coordinate. The analysis of analytic solutions in alternative
gravitational theories has attained much attention in recent years
to study cosmic evolution and acceleration. The direct solution of
the above equations is very difficult due to their highly non-linear
nature. There are two possible ways to solve these equations. One is
to solve them by applying suitable exact or numeric method whereas
2nd is to obtain exact solutions by Noether symmetry technique. This
theory is non-conserved but conserved quantities can be examined by
Noether symmetries. Thus the latter approach appears to be more
interesting and we adopt it in this article.

\section{Noether Symmetries}

Noether symmetries offer an interesting method to establish new
cosmological models and associated structures in modified theories.
This section develops point-like Lagrangian and determines the
corresponding Noether equations for the BT-I universe in the context
of EMSG. This technique yields a unique vector field corresponding
to tangent space. Thus, the vector field acts as a symmetry
generator and gives conserved parameters that are useful to
investigate exact cosmological solutions.

The action (\ref{1}) in canonical form becomes
\begin{equation}\label{11}
\mathcal{S}= \int
\mathrm{L}(\mathrm{a},\dot{\mathrm{a}},\mathrm{b},\dot{\mathrm{b}},
\mathcal{R}, \dot{\mathcal{R}}, \mathbf{T}^{2},
\dot{\mathbf{T}^{2}})dt.
\end{equation}
To obtain point-like Lagrangian, we use the Lagrange multiplier
technique as
\begin{equation}\label{12}
\mathcal{S}= \int
(f+\mathrm{p}-(\mathcal{R}-\bar{\mathcal{R}})\lambda_{1}-
(\mathbf{T}^{2}-\mathbf{\bar{T}}^{2})\lambda_{2})\sqrt{-g}d\mathrm{t},
\end{equation}
where
\begin{equation}\nonumber
\bar{\mathcal{R}}=
2\left(\frac{\ddot{\mathrm{a}}}{\mathrm{a}}+\frac{2\ddot{\mathrm{b}}}
{\mathrm{b}}+\frac{2\dot{\mathrm{a}}\dot{\mathrm{b}}}{\mathrm{a}\mathrm{b}}
+\frac{\dot{\mathrm{b}}^{2}}{\mathrm{b}^{2}}\right), \quad
\bar{\mathbf{T}}^{2}= 3\mathrm{p}^{2}+\rho^{2}, \quad \lambda_{1}=
f_{\mathcal{R}}, \quad \lambda_{2}= f_{\mathbf{T}^{2}}.
\end{equation}
We observe that the action (\ref{1}) for BT-I spacetime is recovered
when $\mathcal{R}-\bar{\mathcal{R}}=0$ and $\mathbf{T}^{2}-
\bar{\mathbf{T}^{2}}=0$. Substituting the above values in
Eq.(\ref{12}), we have
\begin{eqnarray}\nonumber
\mathrm{L}&=&
\mathrm{a}\mathrm{b}^{2}(f+\mathrm{p})-2(\mathrm{a}\dot{\mathrm{b}}^{2}
+2\dot{\mathrm{a}}\dot{\mathrm{b}}\mathrm{b})f_{\mathcal{R}}
+\mathrm{a}\mathrm{b}^{2}\left(3\mathrm{p}^{2}+\rho^{2}\right)f_{\mathbf{T}^{2}}-2
\\\label{13}&\times&
(\dot{\mathrm{a}}\mathrm{b}^{2}+2\mathrm{a}
\mathrm{b}\dot{\mathrm{b}})(\dot{\mathcal{R}}f_{\mathcal{R}\mathcal{R}}
+\dot{\mathbf{T}}^{2}f_{\mathcal{R}\mathbf{T}^{2}})
-\mathrm{a}\mathrm{b}^{2}\left(\mathcal{R}
f_{\mathcal{R}}+\mathbf{T}^{2} f_{\mathbf{T}^{2}}\right).
\end{eqnarray}
It is noteworthy to mention here that when
$\mathrm{a}(\mathrm{t})=\mathrm{b}(\mathrm{t})$, the Lagrangian of
FRW universe is recovered \cite{34}. The above Lagrangian is
difficult due to its highly non-linear nature. Therefore, we need
some additional constraints to solve it. Here we use a physical
condition that the ratio of shear scalar $(\sigma)$ to expansion
scalar $(\Theta)$ is constant, which gives
$\mathrm{a}=\mathrm{b}^{n}$, where $n$ is a real constant and we
consider $n\neq 0,1$ for non-trivial solution.  This condition
suggests that the Hubble cosmic expansion can achieve isotropy when
$\frac{\sigma}{\Theta}$ is constant \cite{34a}. Several researchers
have also used this relation to obtain the cosmological solutions
\cite{34c}. Thus, the Lagrangian (\ref{13}) becomes
\begin{eqnarray}\nonumber
\mathrm{L}&=&
\mathrm{b}^{n+2}(f+\mathrm{p})-\mathrm{b}^{n+2}\left(\mathcal{R}
f_{\mathcal{R}}+\mathbf{T}^{2} f_{\mathbf{T}^{2}}\right)
+\mathrm{b}^{n+2}\left(3\mathrm{p}^{2}+\rho^{2}\right)f_{\mathbf{T}^{2}}
\\\label{14}&-&2(\mathrm{b}^{n}\dot{\mathrm{b}}^{2}+2n\mathrm{b}^{n}
\dot{\mathrm{b}}^{2})f_{\mathcal{R}}-2\mathrm{b}^{n+1}\dot{\mathrm{b}}(n+2)
(\dot{\mathcal{R}}f_{\mathcal{R}\mathcal{R}}+\dot{\mathbf{T}}^{2}
f_{\mathcal{R}\mathbf{T}^{2}}).
\end{eqnarray}
The corresponding Hamiltonian, $\mathcal{H}= \dot{q}^{i}(\partial
\mathrm{L}/\partial \dot{q}^{i})-\mathrm{L}$ is given by
\begin{eqnarray}\nonumber
\mathcal{H}&=&
\mathrm{b}^{n+2}\left(\mathcal{R}f_{\mathcal{R}}+\mathbf{T}^{2}
f_{\mathbf{T}^{2}}\right)-\mathrm{b}^{n+2}\left(3\mathrm{p}^{2}
+\rho^{2}\right)f_{\mathbf{T}^{2}}-\mathrm{b}^{n+2}(f+\mathrm{p})
\\\label{15}&-&2(\mathrm{b}^{n}\dot{\mathrm{b}}^{2}+2n\mathrm{b}^{n}
\dot{\mathrm{b}}^{2})f_{\mathcal{R}}-2\mathrm{b}^{n+1}\dot{\mathrm{b}}
(n+2)\dot{f_{\mathcal{R}}}.
\end{eqnarray}

The symmetry generators of the Lagrangian (\ref{14}) are given by
\begin{equation}\label{16}
\mathrm{K}= \xi(\mathrm{t},\mathrm{b}, \mathcal{R},
\mathbf{T}^{2})\frac{\partial}{\partial
t}+\eta^{i}(\mathrm{t},\mathrm{b}, \mathcal{R},
\mathbf{T}^{2})\frac{\partial}{\partial q^{i}}, \quad i=1,2,3,
\end{equation}
where undetermined coefficients of $\mathrm{K}$ are represented by
$\xi$ and $\eta^{i}=(\alpha, \beta, \gamma)$, respectively. For the
existence of Noether symmetries, the Lagrangian must fulfill the
invariance condition defined as
\begin{equation}\label{17}
\mathrm{K}^{[1]}\mathrm{L}+(\mathfrak{D}\xi)\mathrm{L}=
\mathfrak{D}\psi,
\end{equation}
where the total rate of change, first-order prolongation, and
boundary term are denoted by $\mathfrak{D}$, $\mathrm{K}^{[1]}$ and
$\psi$, respectively. Further, it is expressed as
\begin{equation}\label{18}
\mathrm{K}^{[1]}= \mathrm{K}+\dot{\eta}^{i}\frac{\partial}{\partial
\dot{q}^{i}}, ~~~\mathfrak{D}= \frac{\partial}{\partial
t}+\dot{q}^{i}\frac{\partial}{\partial q^{i}},
\end{equation}
here
$\dot{\eta}^{i}$=$\mathfrak{D}\eta^{i}-\dot{q}^{i}\mathfrak{D}\xi$.
The corresponding conserved quantities are defined as
\begin{equation}\label{19}
\mathcal{I}= \eta^{i}\frac{\partial
\mathrm{L}}{\partial\dot{q}^{i}}-\xi \mathcal{H}-\psi.
\end{equation}
This is also dubbed as the first integral of motion and considered
the most crucial part of Noether symmetries which play a significant
role to determine viable cosmological solutions.

We formulate the following system of partial differential equations
(PDEs) by evaluating and comparing the coefficients of Eq.(\ref{17})
\begin{eqnarray}\label{20}
\xi_{\mathrm{b}}=0, \quad \xi_{\mathcal{R}}=0, \quad
\xi_{\mathbf{T}^{2}}=0,
\end{eqnarray}
\begin{eqnarray}\label{21}
(2+n)\mathrm{b}^{n+1}\alpha_{\mathcal{R}}f_{\mathcal{R}\mathcal{R}}=0,
\quad 4\mathrm{b}^{n+1}\alpha_{\mathrm{t}}f_{\mathcal{R}\mathcal{R}}
+2n\mathrm{b}^{n+1}\alpha_{\mathrm{t}}f_{\mathcal{R}\mathcal{R}}
-\psi_{\mathcal{R}}=0,
\end{eqnarray}
\begin{eqnarray}\label{22}
(2+n)\mathrm{b}^{n+1}\alpha_{\mathbf{T}^{2}}f_{\mathcal{R}\mathbf{T}^{2}}=0,
\quad
4\mathrm{b}^{n+1}\alpha_{\mathrm{t}}f_{\mathcal{R}\mathbf{T}^{2}}
+2n\mathrm{b}^{n+1}\alpha_{\mathrm{t}}f_{\mathcal{R}\mathbf{T}^{2}}
-\psi_{\mathbf{T}^{2}}=0,
\end{eqnarray}
\begin{eqnarray}\label{23}
4\mathrm{b}^{n+1}\alpha_{\mathbf{T}^{2}}f_{\mathcal{R}\mathcal{R}}
+2n\mathrm{b}^{n+1}\alpha_{\mathbf{T}^{2}}f_{\mathcal{R}\mathcal{R}}
+4\mathrm{b}^{n+1}\alpha_{\mathcal{R}}f_{\mathcal{R}\mathbf{T}^{2}}
+2n\mathrm{b}^{n+1}\alpha_{\mathcal{R}}f_{\mathcal{R}\mathbf{T}^{2}}=0,
\end{eqnarray}
\begin{eqnarray}\nonumber
&&4\mathrm{b}^{n}\alpha_{\mathrm{t}}f_{\mathcal{R}}+8n\mathrm{b}
^{n}\alpha_{\mathrm{t}}f_{\mathcal{R}}
+4\mathrm{b}^{n+1}\beta_{\mathrm{t}}f_{\mathcal{R}\mathcal{R}}
+2n\mathrm{b}^{n+1}\beta_{\mathrm{t}}f_{\mathcal{R}\mathcal{R}}
+4\mathrm{b}^{n+1}\gamma_{\mathrm{t}}
\\\label{24}&&
\times
f_{\mathcal{R}\mathbf{T}^{2}}+2n\mathrm{b}^{n+1}\gamma_{\mathrm{t}}
f_{\mathcal{R}\mathbf{T}^{2}}
-\psi_{\mathrm{b}}=0,
\end{eqnarray}
\begin{eqnarray}\nonumber
&&-2n(1+2n)\alpha
f_{\mathcal{R}}-2(1+2n)\xi_{\mathrm{t}}f_{\mathcal{R}}
+2(1+2n)\alpha_{,\mathrm{b}}f_{\mathcal{R}}+\beta(1+2n)
\\\nonumber&&
\times
f_{\mathcal{R}\mathcal{R}}+2\mathrm{b}f_{\mathcal{R}\mathcal{R}}
\beta_{,\mathrm{b}}+\mathrm{b}nf_{\mathcal{R}\mathcal{R}}
\beta_{,\mathrm{b}}+f_{\mathcal{R}\mathbf{T}^{2}}\gamma
+2nf_{\mathcal{R}\mathbf{T}^{2}}\gamma+2\mathrm{b}f_{\mathcal{R}
\mathbf{T}^{2}}\gamma_{,\mathrm{b}}
\\\label{25}&&
+n\mathrm{b}f_{\mathcal{R}\mathbf{T}^{2}}\gamma_{,\mathrm{b}}
-2\mathrm{b}\xi=0,
\end{eqnarray}
\begin{eqnarray}\nonumber
&&-4\mathrm{b}^{n}\alpha_{,\mathbf{T}^{2}}f_{\mathcal{R}}-8n\mathrm{b}
^{n}\alpha_{,\mathbf{T}^{2}}f_{\mathcal{R}}
-4\mathrm{b}^{n+1}\beta_{,\mathbf{T}^{2}}f_{\mathcal{R}\mathcal{R}}
-2n\mathrm{b}^{n+1}\beta_{,\mathbf{T}^{2}}f_{\mathcal{R}\mathcal{R}}
-4
\\\nonumber&&
\times
\alpha\mathrm{b}^{n+1}f_{\mathcal{R}\mathbf{T}^{2}}-6n\mathrm{b}^{n}
\alpha_{,\mathbf{T}^{2}}f_{\mathcal{R}\mathbf{T}^{2}}
-2n^{2}\mathrm{b}^{n}\alpha
f_{\mathcal{R}\mathbf{T}^{2}}+8\mathrm{b}^{1+n}\xi_{,\mathrm{t}}
f_{\mathcal{R}\mathbf{T}^{2}}+4n
\\\nonumber&&
\times
\xi_{,\mathrm{t}}\mathrm{b}^{1+n}f_{\mathcal{R}\mathbf{T}^{2}}
-4\mathrm{b}^{1+n}\alpha_{,\mathrm{b}}f_{\mathcal{R}\mathbf{T}^{2}}
-2n\mathrm{b}^{1+n}
\alpha_{,\mathrm{b}}f_{\mathcal{R}\mathbf{T}^{2}}-4\mathrm{b}^{1+n}
\gamma_{,\mathbf{T}^{2}}f_{\mathcal{R}\mathbf{T}^{2}}
\\\nonumber&&
-2n\gamma_{,\mathbf{T}^{2}}\mathrm{b}^{1+n}
f_{\mathcal{R}\mathbf{T}^{2}}-4\mathrm{b}^{1+n}\beta
f_{\mathcal{R}\mathcal{R}\mathbf{T}^{2}} -2n\mathrm{b}^{1+n}\beta
f_{\mathcal{R}\mathcal{R}\mathbf{T}^{2}} -4\gamma\mathrm{b}^{1+n}
\\\label{26}&&
\times
f_{\mathcal{R}\mathbf{T}^{2}\mathbf{T}^{2}}-2n\mathrm{b}^{1+n}\gamma
f_{\mathcal{R}\mathbf{T}^{2}\mathbf{T}^{2}}=0,
\end{eqnarray}
\begin{eqnarray}\nonumber
&&-4\mathrm{b}^{n}\alpha_{,\mathcal{R}}f_{\mathcal{R}}-8n\mathrm{b}
^{n}\alpha_{,\mathcal{R}}f_{\mathcal{R}} -4\mathrm{b}^{n}\alpha
f_{\mathcal{R}\mathcal{R}} -6n\mathrm{b}^{n}\alpha
f_{\mathcal{R}\mathcal{R}} -2n^{2}\mathrm{b}^{n}\alpha
f_{\mathcal{R}\mathcal{R}}
\\\nonumber&&
+8\mathrm{b}^{1+n}\xi_{,\mathbf{T}^{2}}f_{\mathcal{R}\mathcal{R}}
+4n\mathrm{b}^{1+n}\xi_{\mathrm{t}}f_{\mathcal{R}\mathcal{R}}
-4\mathrm{b}^{1+n}\alpha_{,\mathrm{b}}
f_{\mathcal{R}\mathcal{R}}-2n\mathrm{b}^{1+n}\eta_{,\mathrm{b}}
f_{\mathcal{R}\mathcal{R}}
\\\nonumber&&
-4\mathrm{b}^{1+n}\beta_{,\mathcal{R}}f_{\mathcal{R}\mathcal{R}}
-2n\mathrm{b}^{1+n}
\beta_{,\mathcal{R}}f_{\mathcal{R}\mathcal{R}}-4\mathrm{b}^{1+n}
\gamma_{,\mathcal{R}}f_{\mathcal{R}\mathbf{T}^{2}}-2n\mathrm{b}^{1+n}
\gamma_{,\mathcal{R}}f_{\mathcal{R}\mathbf{T}^{2}}
\\\nonumber&&
-4\mathrm{b}^{1+n}\beta f_{\mathcal{R}\mathcal{R}\mathcal{R}}
-2n\mathrm{b}^{1+n}\beta f_{\mathcal{R}\mathcal{R}\mathcal{R}}
-4\mathrm{b}^{1+n}\gamma f_{\mathcal{R}\mathcal{R}\mathbf{T}^{2}}
-2n\mathrm{b}^{1+n}\gamma f_{\mathcal{R}\mathcal{R}\mathbf{T}^{2}}
\\\label{27}&&
=0,
\end{eqnarray}
\begin{eqnarray}\nonumber
&&\alpha(n+2)f-\alpha(n+2)(\mathcal{R}
f_{\mathcal{R}}-\mathbf{T}^{2}f_{\mathbf{T}^{2}})
+\alpha(n+2)(3\mathrm{p}^{2}+\rho^{2})f_{\mathbf{T}^{2}}
\\\nonumber&&
+(n+2)\alpha\mathrm{p}+\mathrm{b}\alpha((6\mathrm{p}
\mathrm{p}_{\mathrm{b}}
+2\rho\rho_{\mathrm{b}})f_{\mathbf{T}^{2}}
+\mathrm{p}_{\mathrm{b}})+\mathrm{b}\beta(-\mathcal{R}
f_{\mathcal{R}\mathcal{R}}-\mathbf{T}^{2}
f_{\mathcal{R}\mathbf{T}^{2}}
\\\nonumber&&
+(3\mathrm{p}^{2}+\rho^{2})f_{\mathcal{R}\mathbf{T}^{2}})+\mathrm{b}
\gamma(-\mathcal{R}
f_{\mathcal{R}\mathbf{T}^{2}}-\mathbf{T}^{2}f_{\mathbf{T}^{2}\mathbf{T}^{2}}
+(3\mathrm{p}^{2}+\rho^{2})f_{\mathbf{T}^{2}\mathbf{T}^{2}})
\\\nonumber&&
+\mathrm{b}\xi_{,\mathrm{t}}f-\mathrm{b}\xi_{,\mathrm{t}}(\mathcal{R}
f_{\mathcal{R}}-\mathbf{T}^{2}
f_{\mathbf{T}^{2}})
+\mathrm{b}\xi_{,\mathrm{t}}(3\mathrm{p}^{2}+\rho^{2})f_{\mathbf{T}^{2}}
+\mathrm{b}\xi_{,\mathrm{t}}\mathrm{p}-\psi_{,\mathrm{t}}
\\\label{28}&&
=0.
\end{eqnarray}
These equations play a key role to study the cosmic mysteries in the
context of EMSG. We solve the above system of PDEs in the next
section to find some physically viable solutions for different EMSG
models. Furthermore, we choose $n=-0.5$  to evaluate the
cosmological solutions as we are unable to find exact solutions for
other values of $n$.

\section{Exact Anisotropic Solutions}

This section gives a detailed analysis to formulate the symmetry
generators, first integrals of motion and the corresponding viable
solutions through the system of PDEs. Since the above system is very
complicated and highly non-linear, therefore, it is difficult to
obtain analytic solutions without taking any specific EMSG model. We
consider different models which minimize the system complexity and
help to derive exact solutions. In the first two cases, we examine
the cosmic evolution for dust matter.

\subsection*{Case I: Minimal Coupling Model}

This case analyzes the EMSG by considering the particular form of a
generic function, i.e., minimal coupling among geometric and matter
as $f(\mathcal{R},\mathbf{T}^{2})=\mathcal{R}^{m}+\mathbf{T}^{2}$,
where $m$ is an arbitrary constant \cite{9}. Manipulating
(\ref{20})-(\ref{28}), we obtain
\begin{eqnarray}\nonumber
&&\rho=-\frac{\sqrt{3\mathrm{b}^{\frac{3}{2}}c_{2}\left(39c_{1}\mathcal{R}
\mathrm{b}^{\frac{5}{2}}+50c_{3}m\mathcal{R}
\mathrm{b}^{\frac{3}{2}}(m-1)+75c_{2}\mathcal{R}^{m}
\mathrm{b}^{\frac{3}{2}}+75c_{2}c_{4}\right)}}{15c_{2}\mathrm{b}^{\frac{3}{2}}},
\\\nonumber&&
\alpha=c_{2}\mathrm{b}+\frac{F_{2}(\mathrm{t})}{\sqrt{\mathrm{b}}},
\quad \psi=
3m\mathcal{R}^{m-1}\dot{F_{2}}(\mathrm{t})+F_{3}(\mathrm{t})
+\dot{F_{4}}(\mathrm{t})\mathrm{b}^{\frac{3}{2}}, \quad \xi=c_{1},
\\\label{29}&&
\beta=
\frac{1.33\mathcal{R}^{2-m}c_{1}\mathrm{b}}{m(m-1)}+\frac{\mathcal{R}^{2-m}
F_{4}(\mathrm{t})}{2m(m-1)}-\frac{1.5\mathcal{R}c_{2}}{m-1}+\mathcal{R}^{2-m}c_{3},
\quad \gamma =0,
\end{eqnarray}
where $c_{i}$ represents the arbitrary constant. The symmetry
generators become
\begin{eqnarray}\nonumber
&&\mathrm{K}_{1}= \frac{\partial}{\partial
\mathrm{t}}+\frac{1.33\mathcal{R}^{2-m}\mathrm{b}}{m(m-1)}\frac{\partial}{\partial
\mathcal{R}}, \quad \mathrm{K}_{2}=
\mathrm{b}\frac{\partial}{\partial
\mathrm{b}}-\frac{1.5\mathcal{R}}{m-1}\frac{\partial}{\partial
\mathcal{R}}, \quad \mathrm{K}_{3}=
\mathcal{R}^{2-m}\frac{\partial}{\partial \mathcal{R}},
\end{eqnarray}
and the corresponding conserved quantities are
\begin{eqnarray}\nonumber
\mathcal{I}_{1}&=&-3.99\mathrm{b}^{1.5}\dot{\mathrm{b}}
+3m\mathrm{b}^{0.5}\dot{\mathrm{b}}\dot{\mathcal{R}}
\mathcal{R}^{m-2}(m-1)+\mathrm{b}^{1.5}\mathcal{R}^{m}(1-m)
\\\label{30}
&+&\mathrm{b}^{1.5}(3\mathrm{p}^{2}+\rho^{2}+\mathrm{p}),
\\\label{31}
\mathcal{I}_{2}&=&3m\mathrm{b}^{1.5}\dot{\mathcal{R}}
\mathcal{R}^{m-2}(1-m)+4.5m\mathrm{b}^{0.5}\dot{\mathrm{b}}
\mathcal{R}^{m-1}(m-1),
\\\label{32}
\mathcal{I}_{3}&=&3m\mathrm{b}^{0.5}\dot{\mathrm{b}}(1-m).
\end{eqnarray}
The first and second conserved quantities turn out to be much
complicated, hence exact solution of these quantities is very
difficult. However, one can find numerical solution of these
equations by using some specific conditions. The analytic solution
of Eq.(\ref{32}) is given by
\begin{equation}\label{33}
\mathrm{b}(\mathrm{t})=\left(\mathrm{b}_{0}+\frac{\mathcal{I}_{3}\mathrm{t}}{2m(m-1)}
\right)^{\frac{2}{3}},
\end{equation}
where $\mathrm{b}_{0}$ is an integration constant.
\begin{figure}
\epsfig{file=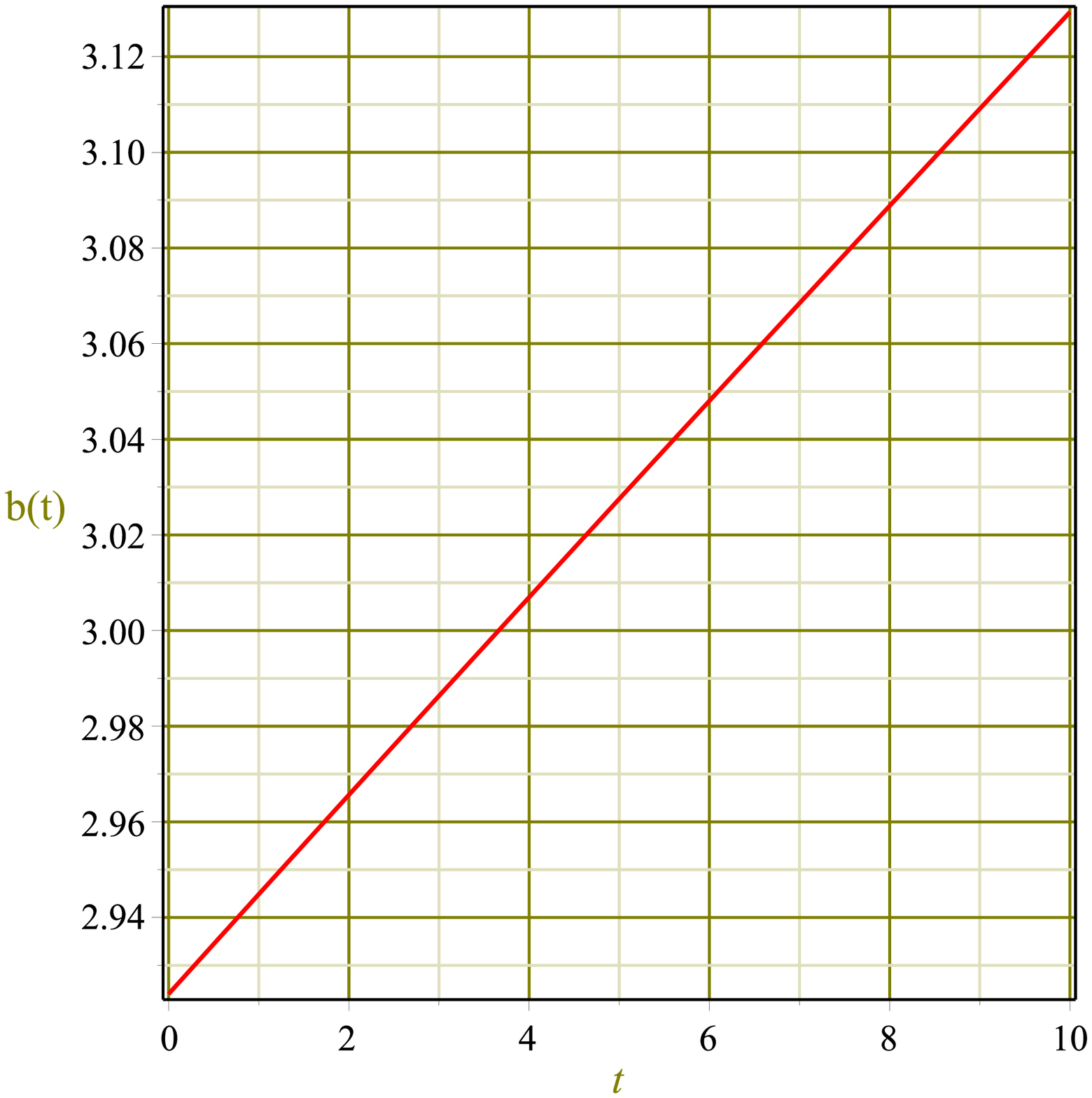,width=.5\linewidth}
\epsfig{file=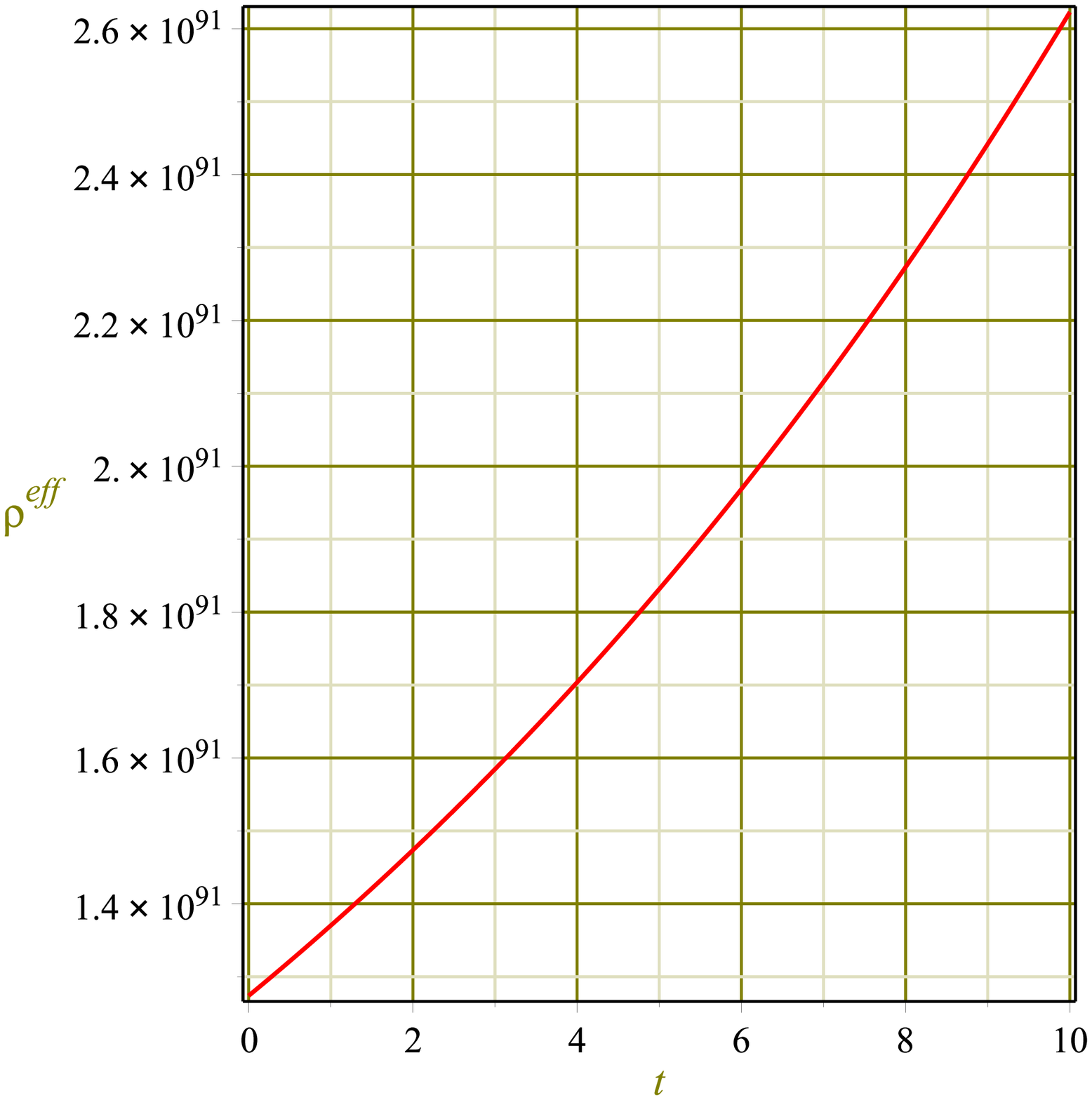,width=.5\linewidth} \caption{Graphs of scale
parameter (left) and energy density (right) corresponding to
$\mathrm{t}$ for minimal coupling model.}
\end{figure}

To analyze this solution, we examine the behavior of some important
physical quantities such as scale factor, EoS parameter and
effective matter variables that play a key role in the study of
current cosmic acceleration. The EoS parameter describes distinct
cosmic eras such as cosmological constant $(\omega=-1)$, phantom
$(\omega<-1)$ and quintessence $(-1<\omega\leq -\frac{1}{3})$ eras.
Here, we consider $m=8$ for our convenience. The graphical
representation of physical quantities is the same for all even
values of $m$ whereas for all odd values, we obtain the reverse
behavior. The corresponding behavior of scale factor and energy
density for the proposed model is shown in Figure \textbf{1}. This
shows that the scale parameter and energy density are positive and
monotonically increasing which determines that the universe is
expanding at a faster rate. Figure \textbf{2} indicates that the
pressure is negatively decreasing and the EoS parameter corresponds
to the quintessence era. Thus, EMSG supports the current cosmic
accelerated expansion due to positively increasing behavior of the
scale parameter/energy density and negatively decreasing behavior of
pressure/EoS parameter.
\begin{figure}
\epsfig{file=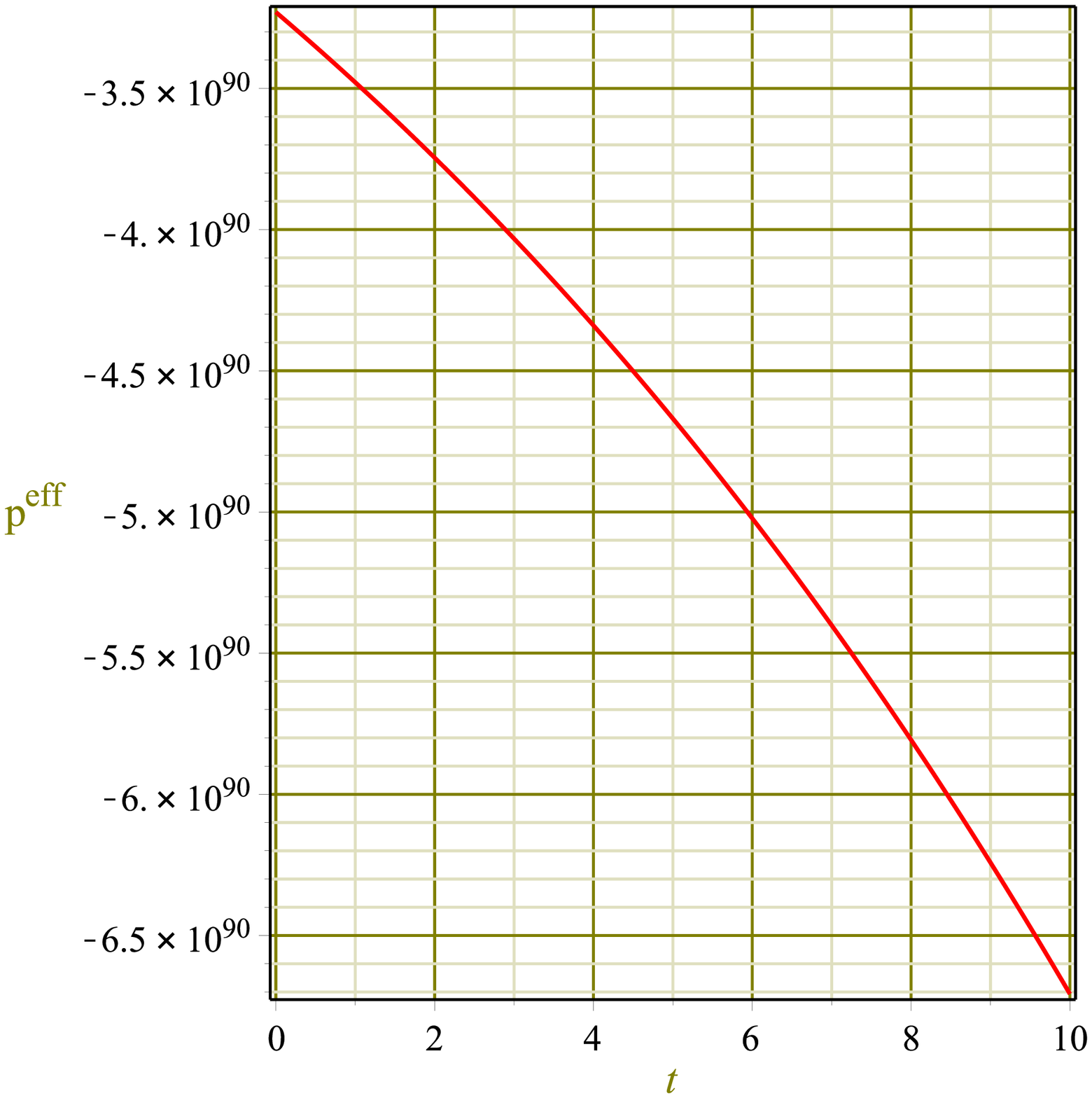,width=.5\linewidth}
\epsfig{file=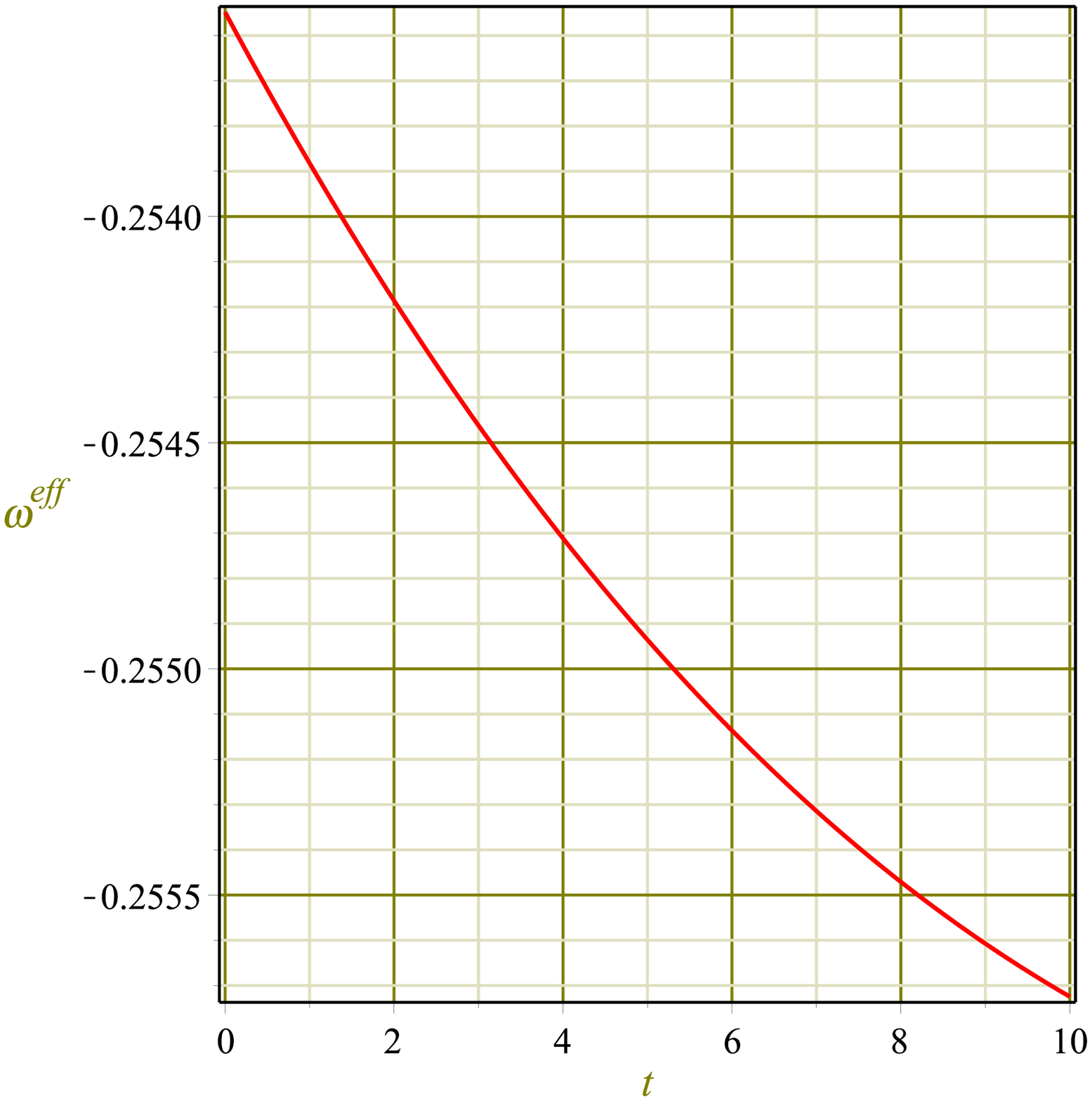,width=.5\linewidth} \caption{Graphs of pressure
(left) and EoS (right) corresponding to $\mathrm{t}$ for minimal
coupling model.}
\end{figure}

\subsection*{Case II: Non-minimal Coupling Model}

This case investigates the EMSG by taking
$f(\mathcal{R},\mathbf{T}^{2})=f_{0}\mathcal{R}\mathbf{T}^{2}$,
where $f_{0}$ is an arbitrary constant \cite{9}. Solving
Eqs.(\ref{20})-(\ref{28}), we obtain
\begin{eqnarray}\nonumber
&&\alpha=c_{2}\mathrm{b}+\frac{F_{2}(\mathrm{t})}{\sqrt{\mathrm{b}}},
\quad \psi=
3f_{0}\mathbf{T}^{2}\dot{F_{2}}(\mathrm{t})+F_{1}(\mathrm{t})
+\dot{F_{4}}(\mathrm{t})\mathrm{b}^{\frac{3}{2}},
\\\nonumber&&
\gamma =-1.5\mathbf{T}^{2}c_{2}+
\frac{1.33c_{3}\mathrm{b}}{f_{0}}+\frac{0.5
F_{4}(\mathrm{t})}{f_{0}}+c_{1}, \quad \xi=c_{3},
\\\label{34}&&
\rho=\frac{\sqrt{3f_{0}c_{2}\mathrm{b}^{\frac{3}{2}}\left(39c_{3}
\mathrm{b}^{\frac{5}{2}}+50c_{1}f_{0}
\mathrm{b}^{\frac{3}{2}}+75c_{2}c_{4}f_{0}\right)}}{15f_{0}
c_{2}\mathrm{b}^{\frac{3}{2}}}, \quad \beta=0.
\end{eqnarray}
The symmetry generators and corresponding conserved parameters are
\begin{eqnarray}\label{35}
\mathrm{K}_{2}&=& \mathrm{b}\frac{\partial}{\partial
\mathrm{b}}-1.5\mathbf{T}^{2}\frac{\partial}{\partial
\mathbf{T}^{2}}, \quad \mathrm{K}_{3}= \frac{\partial}{\partial
t}+\frac{1.33\mathrm{b}}{f_{0}}\frac{\partial}{\partial
\mathbf{T}^{2}},
\\\label{36}
\mathcal{I}_{1}&=&-3f_{0}\mathrm{b}^{0.5}\dot{\mathrm{b}}, \quad
\mathcal{I}_{2}=-3f_{0}\mathrm{b}^{1.5}\dot{\mathbf{T}^{2}}+4.5f_{0}
\mathrm{b}^{0.5}\dot{\mathrm{b}}\mathbf{T}^{2},
\\\label{37}
\mathcal{I}_{3}&=&-3.99\mathrm{b}^{1.5}\dot{\mathrm{b}}
+3f_{0}\mathrm{b}^{0.5}\dot{\mathrm{b}}\dot{\mathbf{T}^{2}}
+\mathrm{b}^{1.5}\mathcal{R}f_{0}\mathbf{T}^{2}
-\mathrm{b}^{1.5}f_{0}\mathcal{R}\mathbf{T}^{2}.
\end{eqnarray}
The exact solution of the first conserved quantity is given by
\begin{equation}\label{38}
\mathrm{b}(\mathrm{t})=\left(\mathrm{b}_{0}+\frac{\mathcal{I}_{1}
\mathrm{t}}{2f_{0}}\right)^{\frac{2}{3}}.
\end{equation}
This equation plays a significant role to analyze the viability of
EMSG. Figure \textbf{3} shows that the energy density is positive
and tends to zero as we go away from the origin whereas scale
parameter is positively increasing which experiences the accelerated
expansion of the universe. The negative behavior of the pressure as
well as EoS parameter (Figure \textbf{4}) ensure the presence of DE,
which is an important candidate for cosmic acceleration.
\begin{figure}
\epsfig{file=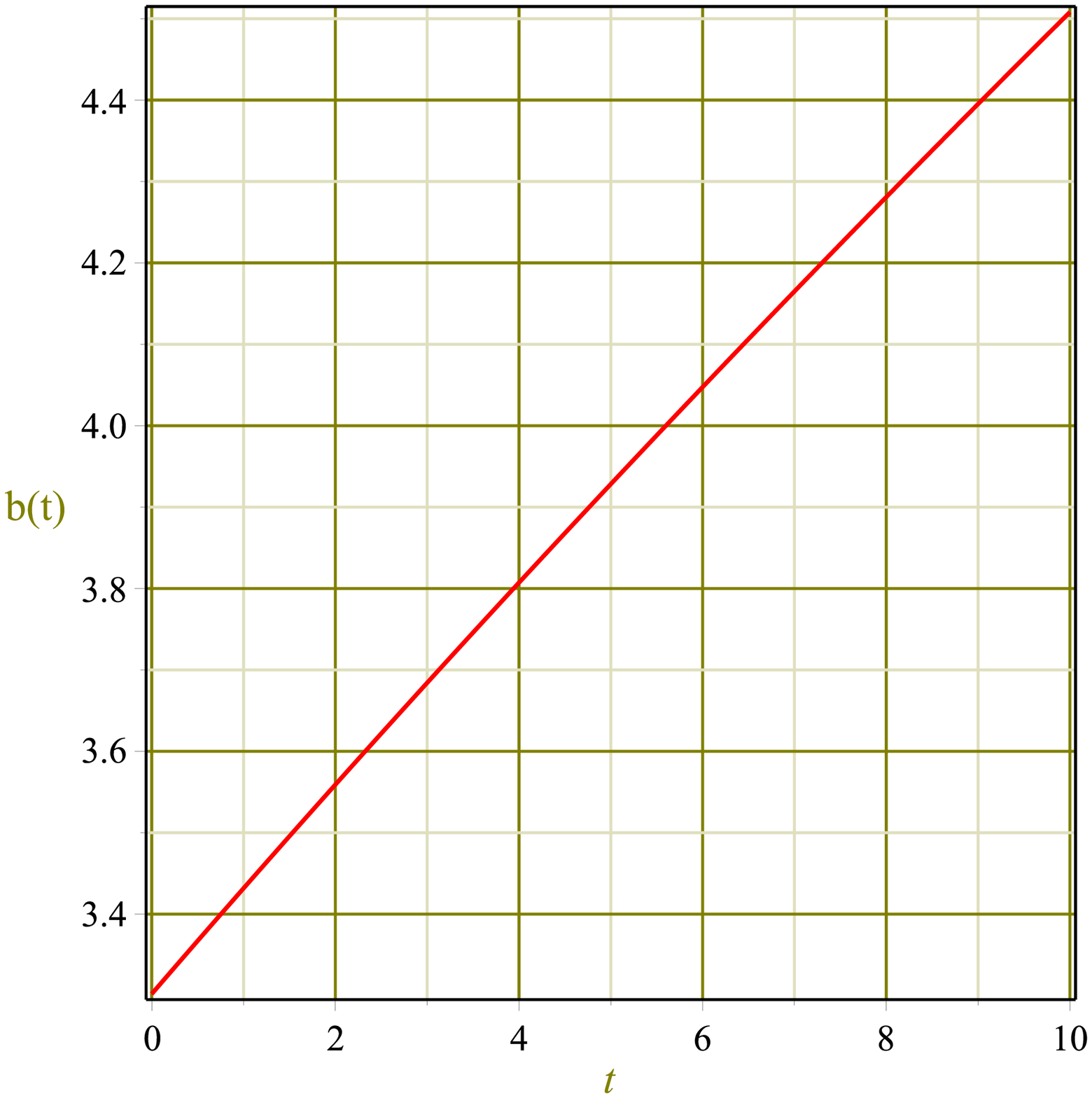,width=.5\linewidth}
\epsfig{file=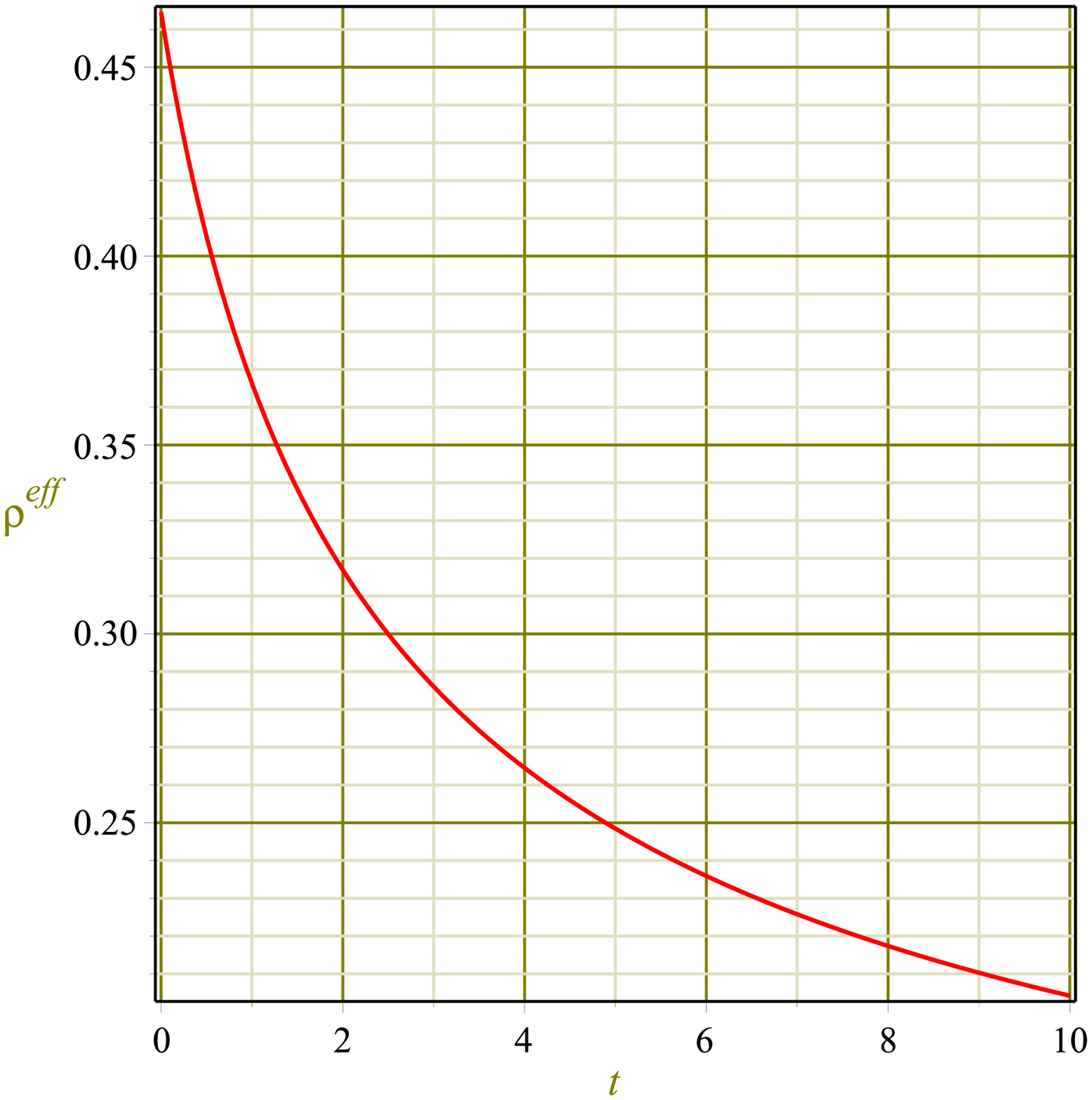,width=.5\linewidth} \caption{Graphs of scale
parameter (left) and energy density (right) corresponding to
$\mathrm{t}$ for non-minimal coupling model.}
\end{figure}
\begin{figure}
\epsfig{file=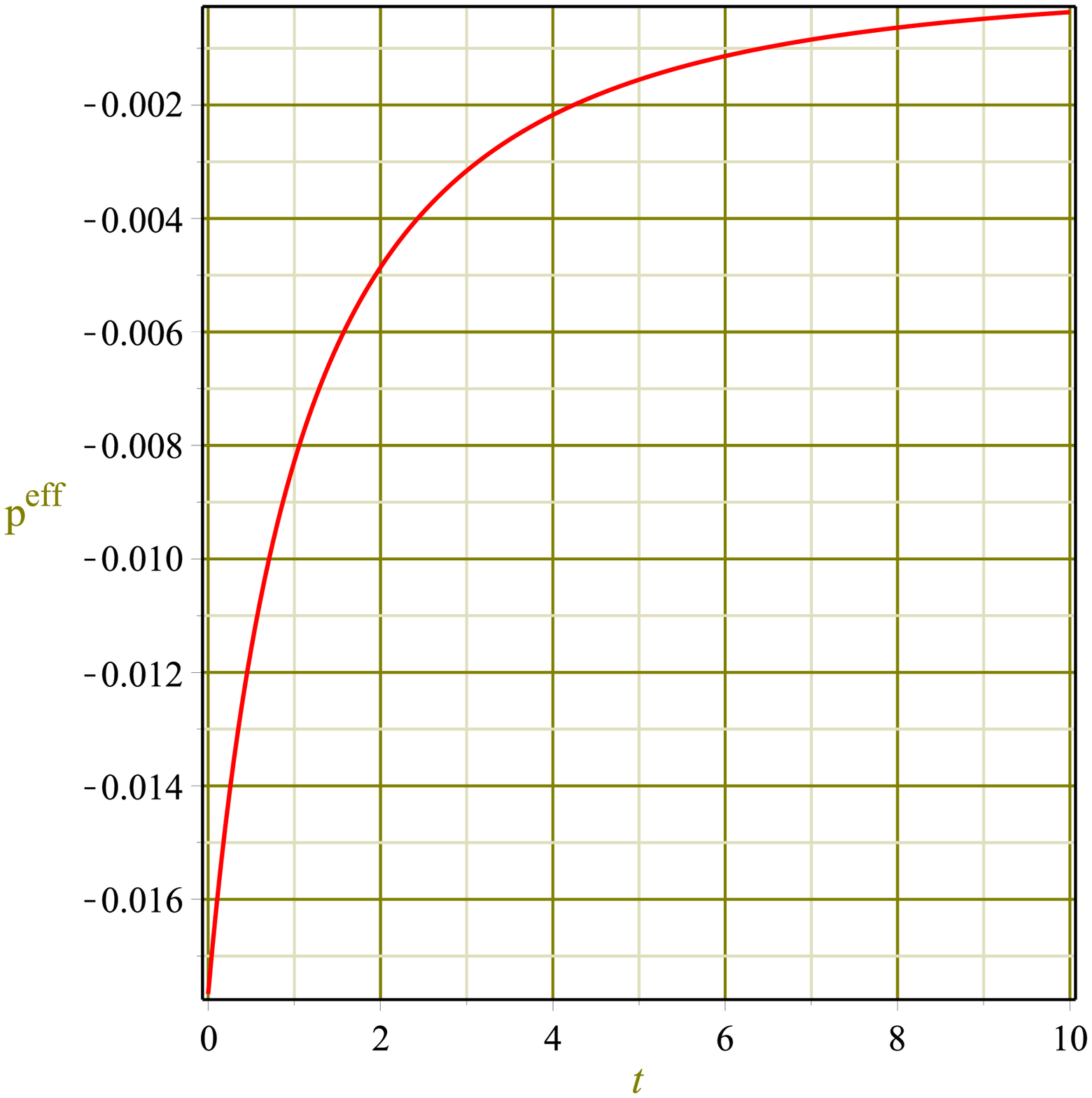,width=.5\linewidth}
\epsfig{file=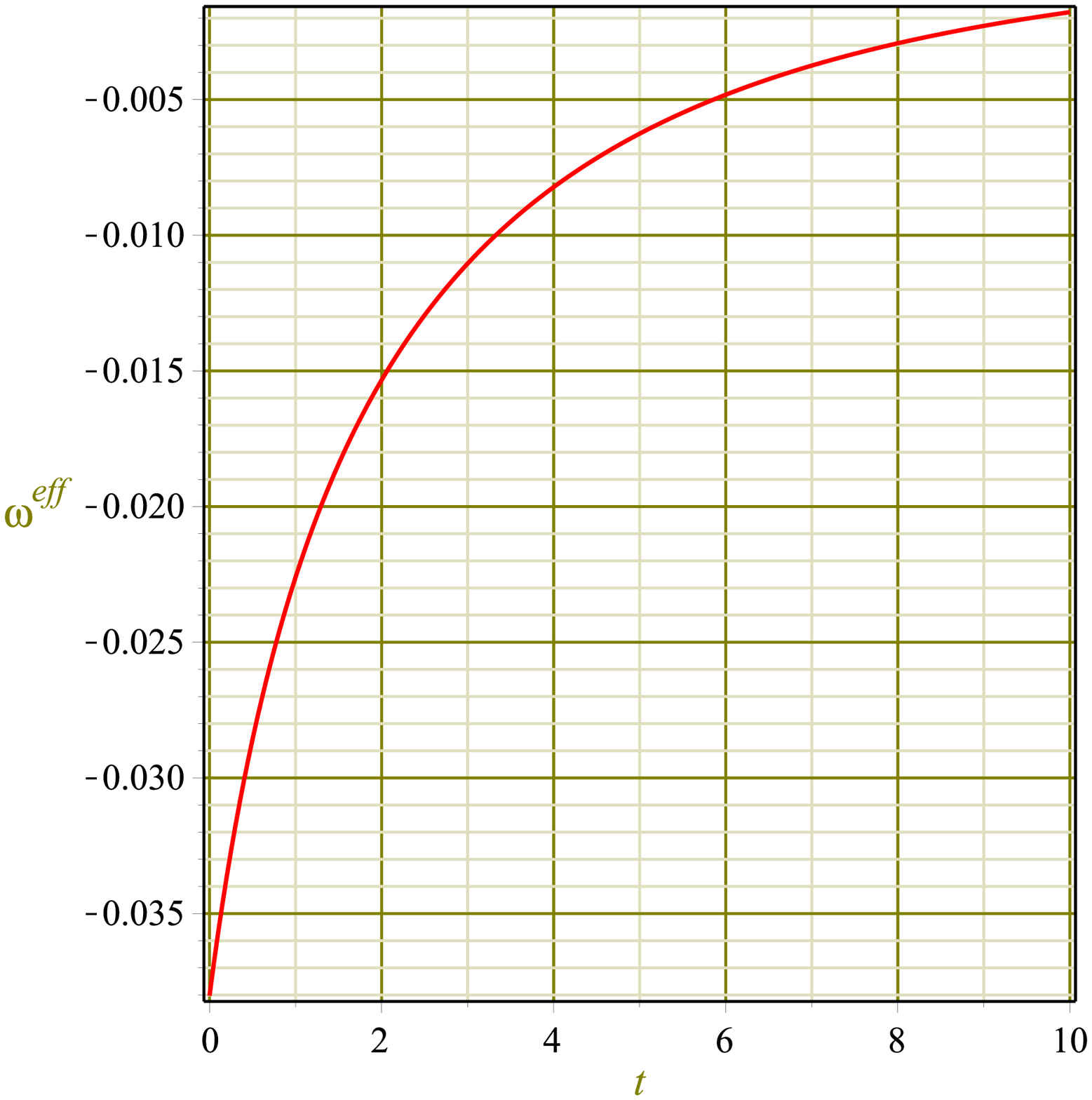,width=.5\linewidth} \caption{Graphs of pressure
(left) and EoS parameter (right) corresponding to $\mathrm{t}$ for
non-minimal coupling model.}
\end{figure}

\subsection*{Case III: $f(\mathcal{R},\mathbf{T}^{2})=f(\mathcal{R})$}
\begin{figure}
\epsfig{file=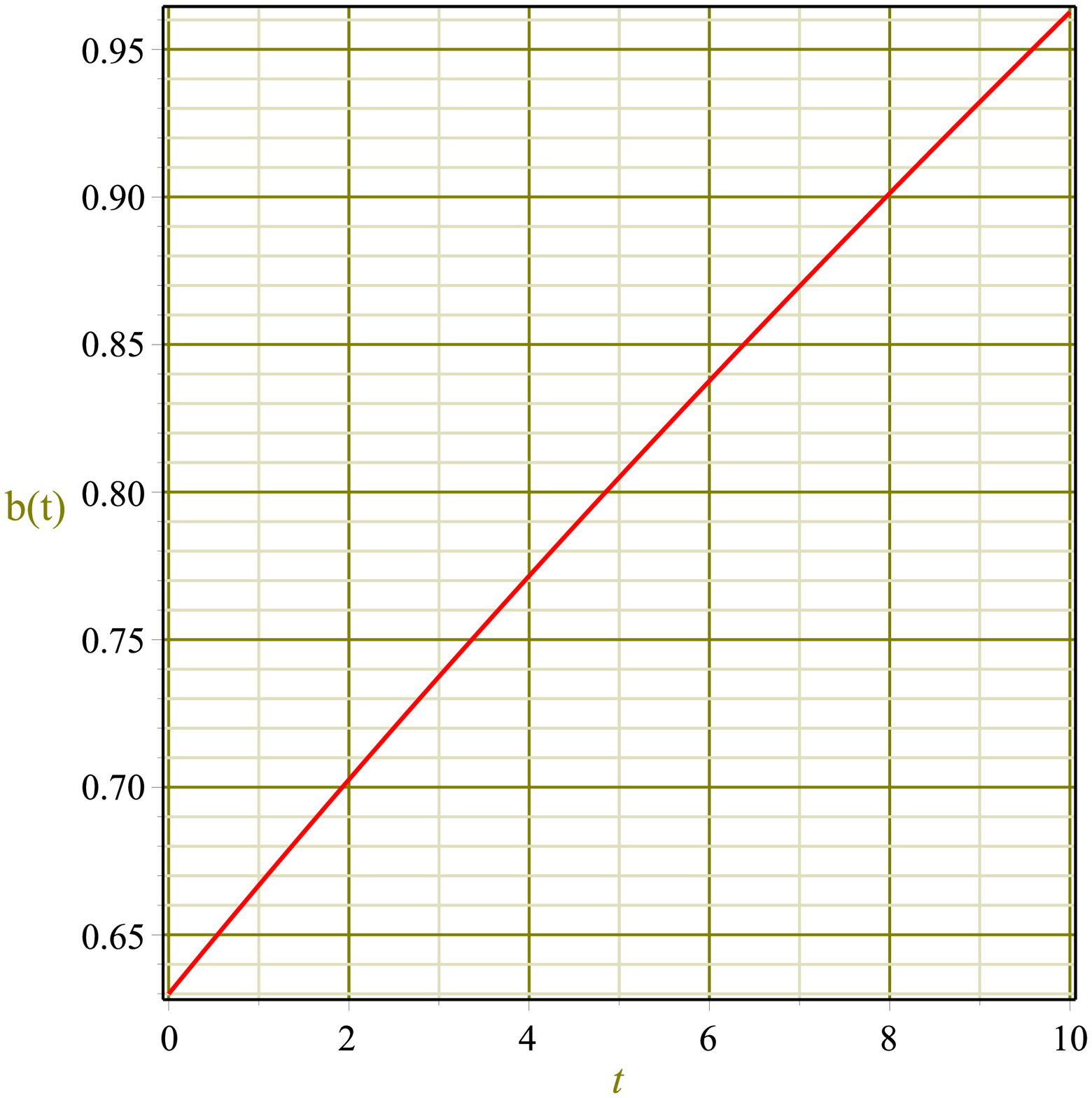,width=.5\linewidth}
\epsfig{file=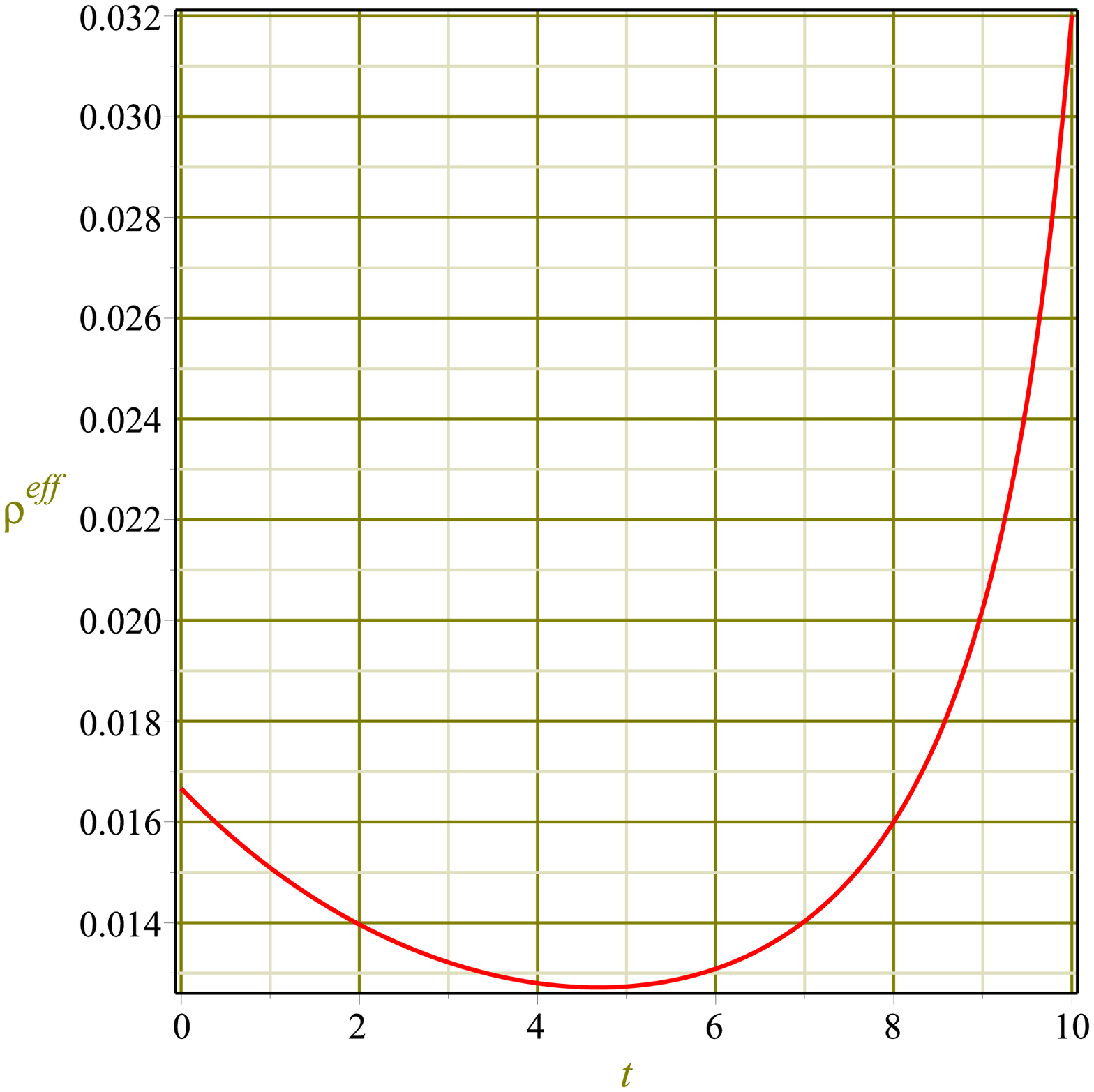,width=.5\linewidth} \caption{Graphs of scale
parameter (left) and energy density (right) corresponding to
$\mathrm{t}$ for $f(\mathcal{R})$ model.}
\end{figure}

Here, we choose
$f(\mathcal{R},\mathbf{T}^{2})=f_{0}\mathcal{R}^{\frac{3}{2}}$ to
re-analyze the $f(\mathcal{R})$ theory. This model has already been
used in various papers \cite{32} which provides viable solutions.
Manipulating (\ref{20})-(\ref{28}), we have
\begin{eqnarray}\nonumber
&&\alpha=
c_{2}\mathrm{b}+\frac{F_{2}(\mathrm{t})}{\sqrt{\mathrm{b}}}, \quad
\psi=
\frac{9}{2}\sqrt{\mathcal{R}}f_{0}\dot{F_{2}}(t)+F_{3}(\mathrm{t})
+\dot{F_{4}}(\mathrm{t})
\mathrm{b}^{\frac{3}{2}}, \quad \xi=c_{1},
\\\nonumber&&
\beta=
\frac{1.78c_{1}\sqrt{\mathcal{R}}\mathrm{b}}{f_{0}}+\frac{0.67F_{4}(\mathrm{t})
\sqrt{\mathcal{R}}}{f_{0}}+(c_{3}-3\sqrt{\mathcal{R}}c_{2})\mathcal{R}^{0.5},
\quad \rho=0,
\\\label{39}&&
\mathrm{p}=\mathrm{b}^{\frac{-3}{2}}c_{4}-f_{0}\mathcal{R}^{\frac{3}{2}}
+\frac{0.534\mathrm{b}\mathcal{R}c_{1}}{c_{2}}
+\frac{0.5\mathcal{R}f_{0}c_{3}}{c_{2}}, \quad \gamma=0.
\end{eqnarray}
The symmetry generators can be determined as
\begin{eqnarray}\nonumber
&&\mathrm{K}_{1}= \frac{\partial}{\partial
t}+\frac{1.78\mathrm{b}\sqrt{\mathcal{R}}}{f_{0}}\frac{\partial}{\partial
\mathcal{R}}, \quad \mathrm{K}_{2}=
\mathrm{b}\frac{\partial}{\partial
\mathrm{b}}-3\mathcal{R}\frac{\partial}{\partial \mathcal{R}}, \quad
\mathrm{K}_{3}= \sqrt{\mathcal{R}}\frac{\partial}{\partial
\mathcal{R}},
\end{eqnarray}
and corresponding conserved quantities become
\begin{eqnarray}\nonumber
\mathcal{I}_{1}&=&\mathrm{b}^{1.5}f-5.340\mathrm{b}^{1.5}\dot{\mathrm{b}}
\sqrt{\mathcal{R}}f_{0}^{-1}f_{\mathcal{R}\mathcal{R}}
+3\mathrm{b}^{0.5}\dot{\mathrm{b}}\dot{\mathcal{R}}
f_{\mathcal{R}\mathcal{R}}+\mathrm{b}^{1.5}\mathrm{p}
\\\label{40}
&-&\mathrm{b}^{1.5}\mathcal{R}f_{\mathcal{R}}-\mathrm{b}^{1.5}
\mathbf{T}^{2}f_{\mathbf{T}^{2}}
+3\mathrm{b}^{1.5}f_{\mathbf{T}^{2}}\mathrm{p}^{2}+3\mathrm{b}^{0.5}
\dot{\mathrm{b}}\dot{\mathbf{T}^{2}}f_{\mathcal{R}\mathbf{T}^{2}},
\\\label{41}
\mathcal{I}_{2}&=&
-3\mathrm{b}^{1.5}\dot{\mathcal{R}}f_{\mathcal{R}\mathcal{R}}
-3\mathrm{b}^{1.5}\dot{\mathbf{T}^{2}}f_{\mathcal{R}\mathbf{T}^{2}}
+9\mathrm{b}^{0.5}\dot{\mathrm{b}}\mathcal{R}f_{\mathcal{R}\mathcal{R}},
\\\label{42}
\mathcal{I}_{3}&=&
-3\mathrm{b}^{0.5}\dot{\mathrm{b}}\mathcal{R}^{0.5}f_{\mathcal{R}\mathcal{R}}.
\end{eqnarray}
We are unable to derive an exact solution from the first and second
conserved quantities $(\mathcal{I}_{1},\mathcal{I}_{2})$ due to
their complicated nature. Therefore, we use
\begin{equation}\label{43}
\mathcal{I}_{3}+3\mathrm{b}^{0.5}\dot{\mathrm{b}}\mathcal{R}^{0.5}
f_{\mathcal{R}\mathcal{R}}=0,
\end{equation}
where solution is
\begin{equation}\label{44}
\mathrm{b}(\mathrm{t})=\left(\mathrm{b}_{0}+\frac{2\mathcal{I}_{3}
\mathrm{t}}{3f_{0}}\right)^{\frac{2}{3}}.
\end{equation}
Figure \textbf{5} indicates the behavior of energy density as
positively decreasing in the initial state but with the passage of
time it shows increasing behavior and $\mathrm{b}(\mathrm{t})$ as
positively increasing which represents the current cosmic
accelerated expansion. Figures \textbf{6} shows that pressure is
negative and has decreasing trend whereas the $\omega$ corresponds
to phantom era which represents the rapid cosmic expansion.
\begin{figure}
\epsfig{file=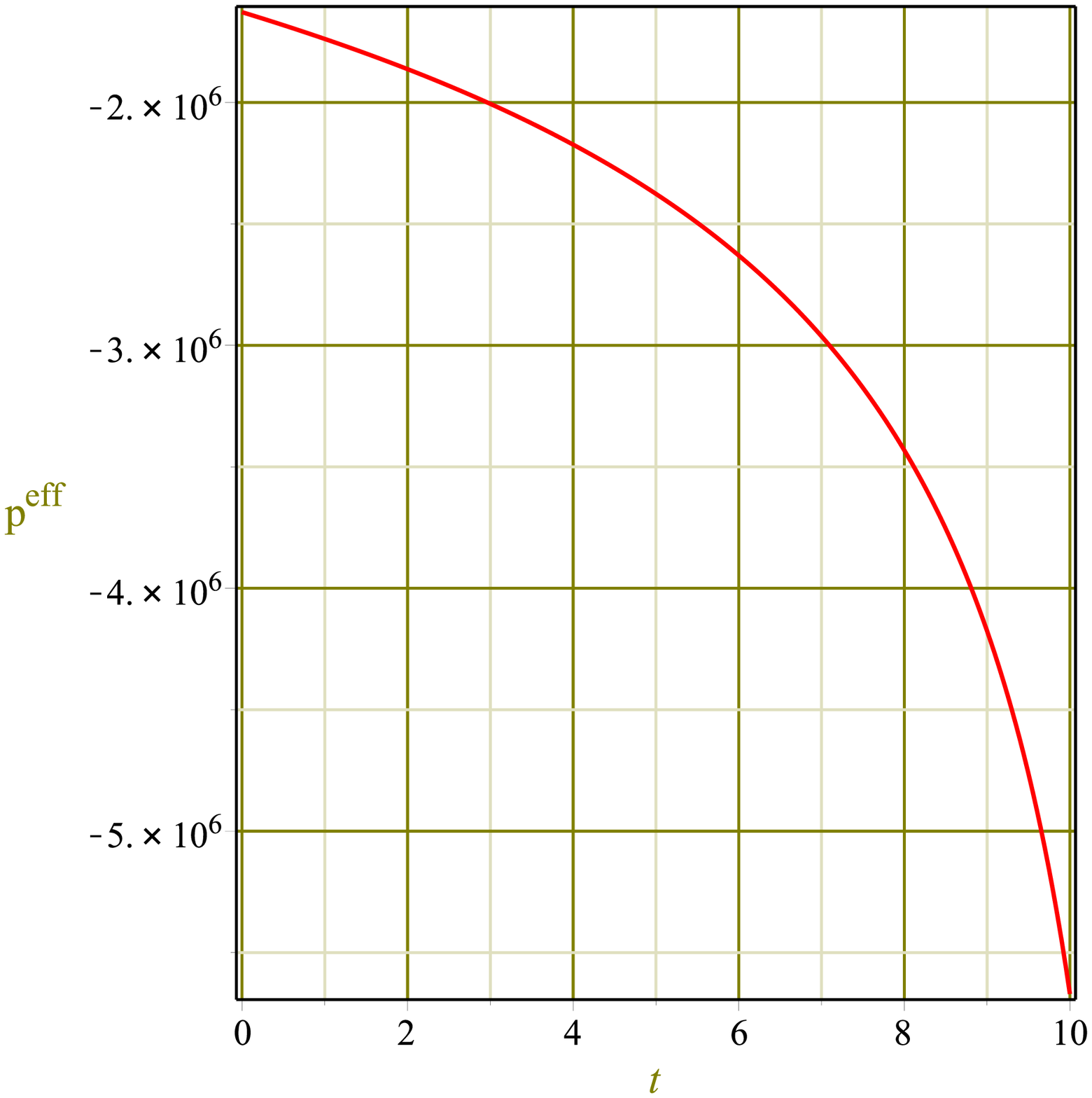,width=.5\linewidth}
\epsfig{file=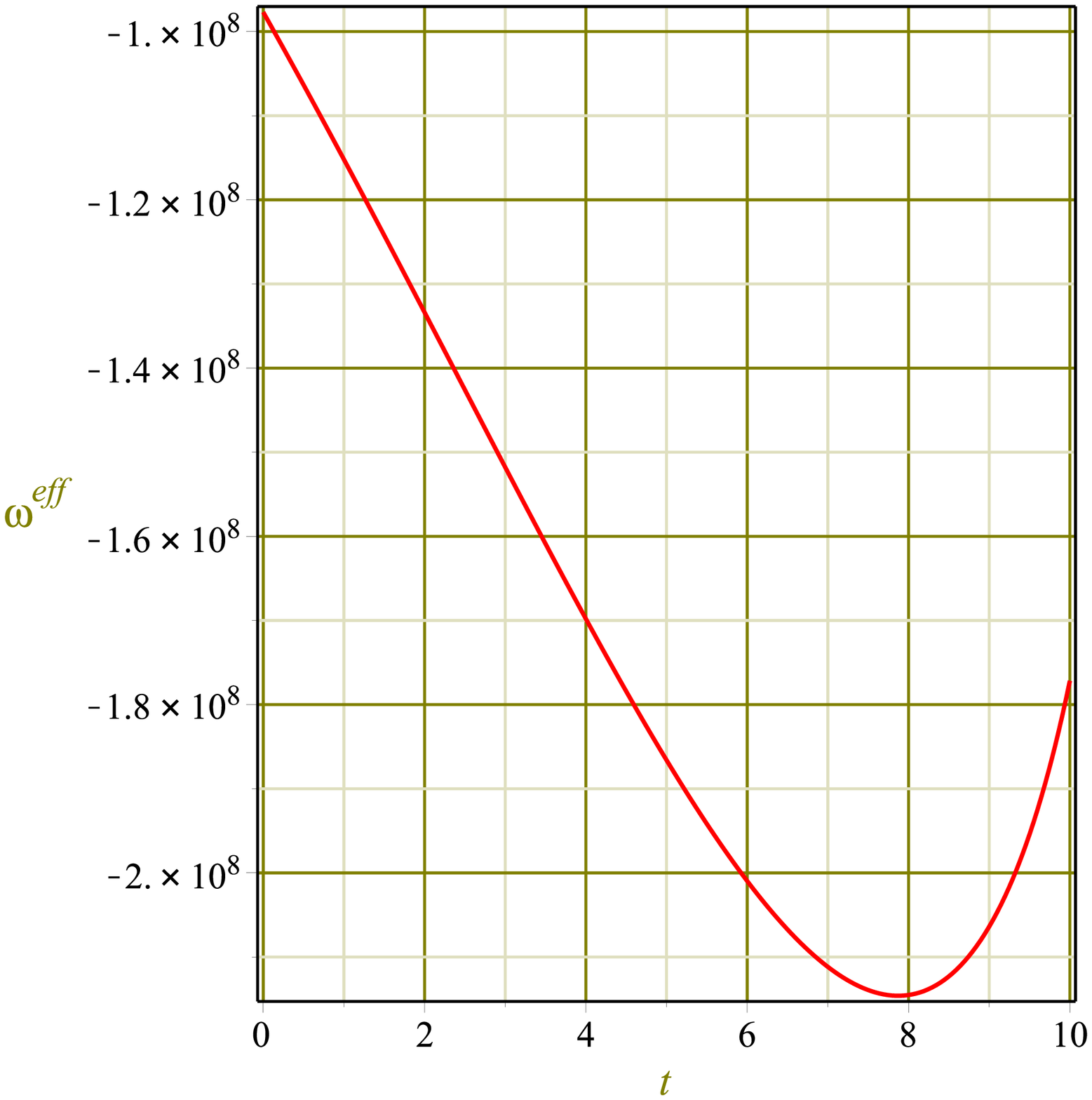,width=.5\linewidth} \caption{Graphs of pressure
(left) and EoS parameter (right) corresponding to $\mathrm{t}$ for
$f(\mathcal{R})$ model.}
\end{figure}

\section{Final Remarks}

In this paper, we have investigated exact cosmological solutions of
BT-I spacetime through Noether symmetry technique in the background
of EMSG. These symmetries provide solutions of the physical system
and their existence yield some suitable conditions to select the
cosmological models based on current observations \cite{37}. The
Lagrangian reduces the system's complexity and helps to derive exact
solutions. We have developed the Lagrangian in the context of EMSG
and formulated conserved parameters to analyze analytic solutions of
the field equations. The exact solutions of Noether equations have
been examined for three different types of functions. The obtained
results are summarized as follows.
\begin{itemize}
\item In the first case, we have formulated three non-zero symmetry generators and
corresponding conserved parameters. The first two conserved
parameters are quite complicated and it is too difficult to obtain
exact solutions from these conserved quantities. The third conserved
quantity provides the exact solution which is examined by analyzing
the behavior of different cosmological parameters. The scale factor
and energy density determine that the universe is in the accelerated
expansion phase (Figure \textbf{1}). The pressure and EoS parameters
show negative behavior which determines the quintessence era
(Figures \textbf{2}).
\item For the second case, the resulting solution indicates that the energy density is
positive and the scale factor shows that our current universe is in
the accelerated expansion phase (Figure \textbf{3}). The negative
behavior of the pressure describes the existence of DE whereas the
EoS parameter manifests the cosmic acceleration (Figures
\textbf{4}).
\item In the last case, the exact solution provides decreasing
trend of the energy density initially but shows increasing behavior as
we go away from the origin while the scale parameter is positive and
has increasing behavior (Figure \textbf{5}). The pressure is
negatively decreasing whereas the EoS parameter corresponds to the
phantom era which shows the rapid expansion of the universe (Figure
\textbf{6}).
\end{itemize}
It is worthwhile to mention here that for $n=1$ all the results
reduce to \cite{34}. In other modified theories of gravity, the
results are consistent corresponding to particular model \cite{39}.
We conclude that all proposed models of
$f(\mathcal{R},\mathbf{T}^{2})$ theory supports the current cosmic
accelerated expansion.


\vspace{0.5cm}

\begin{thebibliography}{55}

\bibitem{1} Perlmutter, S. et al.: Bull. Am. Astron. Soc.
\textbf{29}(1997)1351; Filippenko, A.V. and Riess, A.G.: Phys. Rep.
\textbf{307}(1998)31; Tegmark, M. et al.: Phys. Rev. D
\textbf{69}(2004)103501; Spergel, D.N. et al.: Astrophys. J. Suppl.
\textbf{170}(2007)377.

\bibitem{2} Carroll, S.M.: Living Rev. Relativ. \textbf{4}(2001)1; Sirivastava, S.K.:
\textit{General Relativity and Cosmology} (Prentice Hall of India,
2008).

\bibitem{3} Cognola, G. et al.: Phys. Rev. D \textbf{77}(2008)046009;
Felice, A.D. and Tsujikawa, S.R.: Living Rev. Relativ.
\textbf{13}(2010)3; Nojiri, S. and Odintsov, S.D.: Phys. Rep.
\textbf{505}(2011)59; Bamba, K. et al.: Astrophys. Space Sci.
\textbf{342}(2012)155.

\bibitem{4} Harko, T., Koivisto, T.S. and Lobo, F.S.N.: Mod. Phys. Lett. A
\textbf{26}(2011)1467.

\bibitem{5} Haghani, Z. et al.: Phys. Rev. D \textbf{88}(2013)044023.

\bibitem{6} Katirci, N. and Kavuk, M.: Eur. Phys. J. Plus \textbf{129}(2014)163.

\bibitem{7} Board, C.V.R. and Barrow, J.D.: Phys. Rev. D \textbf{96}(2017)123517.

\bibitem{8} Nari, N. and Roshan, M.: Phys. Rev. D
\textbf{98}(2018)024031;  Akarsu, O. et al.: Phys. Rev. D
\textbf{97}(2018)124017.

\bibitem{9} Bahamonde, S., Marciu, M. and Rudra, P.: Phys. Rev. D
\textbf{100}(2019)083511.

\bibitem{10} Barbar, A.H., Awad, A.M. and AlFiky, M.T.: Phys. Rev. D
\textbf{101}(2020)044058.

\bibitem{11} Singh, K.N. et al.: Phys. Dark Universe \textbf{31}(2021)100774;
Rudra, P. and Pourhassan, B.: arXiv:2008.11034v1.

\bibitem{12} Sharif, M. and Gul, M.Z.: Int. J. Mod. Phys.
A \textbf{36}(2021)2150004; Chin. J. Phys. \textbf{71}(2021)365;
Universe \textbf{07}(2021)154.

\bibitem{13} Barrow, J.D. and Turner, M.S.: Nature \textbf{292}(1982)35; Demianski,
M.: Nature \textbf{307}(1984)140.

\bibitem{14} Akarsu, O. and Kilinc, C.B.: Astrophys. Space Sci. \textbf{326}(2010)315.

\bibitem{15} Yadav, A.K. and Saha, B.: Astrophys. Space Sci. \textbf{337}(2012)759.

\bibitem{16} Adhav, K.S.: Astrophys. Space Sci. \textbf{339}(2012)365.

\bibitem{18} Shamir, M.F.: Eur. Phys. J. C \textbf{75}(2015)8.

\bibitem{19} Sharif, M. and Jabbar, S.: Commun. Theor. Phys. \textbf{63}(2015)168.

\bibitem{21} Demianski, M. et al.: Phys. Rev. D
\textbf{46}(1992)1391.

\bibitem{22} Hanc, J., Tuleja, S. and Hancova, M.: Am. J. Phys.
\textbf{72}(2004)428.

\bibitem{23} Capozziello, S., Marmo, G. and Rubano,
C.P.: Int. J. Mod. Phys. D \textbf{6}(1997)491; Camci,  U.: Eur.
Phys. J. C \textbf{74}(2014)3201; ibid. J. Cosmol. Astropart. Phys.
\textbf{2014}(2014)2.

\bibitem{24} Capozziello, S., Stabile, A. and Troisi, A.:  Class. Quantum Grav.
\textbf{24}(2007)2153; ibid. \textbf{25}(2008)085004;
\textbf{27}(2010)165008.

\bibitem{25} Shamir, M.F., Jhangeer, A. and Bhatti, A.A: Chin. Phys. Lett.
\textbf{29}(2012)080402.

\bibitem{26} Kucukakca, Y.: Eur. Phys. J. C \textbf{73}(2013)2327.

\bibitem{27} Sharif, M. and Waheed, S.: J. Cosmol. Astropart. Phys. \textbf{02}(2013)043.

\bibitem{27a} Sharif, M. and Shafique, I.: Phys. Rev. D \textbf{90}(2014)084033.

\bibitem{28} Momeni, D., Myrzakulov, R. and Gudekli, E.: Int. J. Geom. Methods Mod.
Phys. \textbf{12}(2015)1550101.

\bibitem{29} Sharif, M. and Fatima, H.I.: J. Exp. Theor. Phys. \textbf{122}(2016)104.

\bibitem{30} Shamir, M.F. and Ahmad, M.: Eur. Phys. J. C \textbf{77}(2017)55;
ibid. Mod. Phys. Lett. A \textbf{32}(2017)1750086.

\bibitem{32} Bahamonde, S., Bamba, K. and Camci, U.: J. Cosmol. Astropart. Phys. \textbf{02}(2019)016.

\bibitem{33} Bahamonde, S., Camci, U. and Capozziello, S.: Class. Quantum Grav. \textbf{36}
(2019)065013.

\bibitem{34} Sharif, M. and Gul, M.Z.: Phys. Scr.
\textbf{96}(2020)025002.

\bibitem{35} Sharif, M. and Gul, M.Z.: Eur. Phys. J. Plus
\textbf{136}(2021)503; Adv. Astron. \textbf{2021}(2021)6663502.

\bibitem{34a} Kantowski, R. and Sachs, R.K.: J. Math. Phys.
\textbf{7}(1966)443.

\bibitem{34c} Xing-Xiang, W.: Chin. Phys. Lett. \textbf{22}(2005)29;
Bali, R. and Kumawat, P.: Phys. Lett. B \textbf{665}(2008)332;
Sharif, M. and Zubair,  M.: Astrophys. Space Sci.
\textbf{330}(2010)399

\bibitem{37} Capozziello S, De Laurentis, M. and Odintsov, S.D.: Eur. Phys. J. C
\textbf{72}(2012)2068.

\bibitem{39} Sharif,
M. and Nawazish, I.: Gen. Relativ. Gravit. \textbf{49}(2017)76;
Shamir, M.F. and Kanwal, F.: Eur. Phys. J. C \textbf{77}(2017)1;
Malik, A., Shamir, M.F. and Hussain, I.: Int. J. Geom. Methods Mod.
Phys. \textbf{17}(2020)2050163.

\end{thebibliography}
\end{document}